\documentclass[aps,prl,twocolumn,showpacs,superscriptaddress,footinbib,preprintnumbers]{revtex4-2} 
\usepackage{graphicx}  % needed for figures
\usepackage{dcolumn}   % needed for some tables
\usepackage{bm}        % for math
\usepackage{amssymb}   % for math
\usepackage{amsmath}
\usepackage{comment}
\usepackage{graphicx}
\usepackage{color}
\usepackage{slashed}
\usepackage{hyperref}
\usepackage[dvipsnames]{xcolor}
\usepackage{wasysym}% comment out, only used for a smiley
\usepackage[braket, qm]{qcircuit}
\usepackage{braket}

\definecolor{mygreen}{rgb}{0,0.5,0}
\definecolor{myblue}{rgb}{0,0,0.75}
\definecolor{mymagenta}{cmyk}{0,1,0,0.12}
\definecolor{mygray}{rgb}{0.5,0.5,0.5}

\newcommand{\Eq}[1]{Eq.~(\ref{#1})}

\newcommand{\Eqs}[2]{Eqs.~(\ref{#1}-\ref{#2})}
\newcommand{\Fig}[1]{Fig.~\ref{#1}}

\newcommand{\ua}{\mathord{\uparrow}}
\newcommand{\da}{\mathord{\downarrow}}

\usepackage{umoline}

\begin{document}
\preprint{UMD-PP-021-04}

\title{Thermalization of Gauge Theories from their Entanglement Spectrum}% with topological order}

\author{Niklas Mueller}
\email{niklasmu@umd.edu}
\affiliation{Maryland Center for Fundamental Physics and Department of Physics, 
University of Maryland, College Park, MD 20742, USA.}
\affiliation{Joint Quantum Institute, NIST/University of Maryland, College Park, Maryland 20742, USA. }

\author{Torsten V. Zache}
\affiliation{Center for Quantum Physics, University of Innsbruck, 6020 Innsbruck, Austria}
\affiliation{Institute for Quantum Optics and Quantum Information of the Austrian Academy of Sciences, 6020 Innsbruck, Austria}

\author{Robert Ott}
\affiliation{Heidelberg University, Institut f\"{u}r Theoretische Physik, Philosophenweg 16,
69120 Heidelberg, Germany}

\date{\today}

\begin{abstract} 
Using dual theories embedded into a larger unphysical Hilbert space along entanglement cuts, 
we study the Entanglement Structure of $\mathbf{Z}_2$ lattice gauge theory in $(2+1)$ spacetime dimensions.
We demonstrate Li and Haldane's conjecture, and show consistency of the Entanglement Hamiltonian with the Bisognano-Wichmann theorem.
Studying non-equilibrium dynamics after a quench, we provide an extensive description of thermalization in $\mathbf{Z}_2$ gauge theory which proceeds in a characteristic sequence: 
Maximization of the Schmidt rank
 and spreading of level repulsion at early times, self-similar evolution with scaling coefficients $\alpha = 0.8 \pm 0.2$ and $ \beta = 0.0 \pm 0.1$ at intermediate times, and finally thermal saturation of the von Neumann entropy.
\end{abstract}
\maketitle

\textit{Introduction.} Understanding thermalization of isolated quantum systems is an outstanding challenge in many fields, from atomic gases at ultra-cold temperatures~\cite{eisert2015quantum,schachenmayer2015thermalization,eigen2018universal}, condensed matter physics~\cite{nandkishore2015many,gornyi2005interacting,oganesyan2007localization,borgonovi2016quantum}, cosmology~\cite{micha2004turbulent}, to high energy and nuclear physics~\cite{baier2001bottom,berges2002controlled,arrizabalaga2005equilibration,balasubramanian2011holographic,grozdanov2016strong,berges2020thermalization}. %and quantum information science~\cite{kitaev2003fault}. 
Much progress has been made  in various systems based on the Eigenstate Thermalization Hypothesis (ETH)~\cite{deutsch1991quantum,srednicki1994chaos,d2016quantum}, but not much is known for gauge theories, i.e. systems
with an extensive number of local constraints.
 
Entanglement structure, more precisely the Entanglement Spectrum (ES), first 
suggested by Li and Haldane as an indicator of topological order (TO) in  fractional quantum hall effect states~\cite{li2008entanglement}, has recently become the subject of multiple such studies
\cite{deutsch2013microscopic,yang2015two,khemani2014eigenstate,geraedts2016many,turner2018weak,zhu2020entanglement,Sels2021thermalization,atas2013distribution,giraud2020probing}.
Their extension to lattice gauge theories (LGTs) is ambiguous because gauge invariance allows no local tensor product structure of the physical Hilbert space (HS). This issue has been addressed
in recent years~\cite{buividovich2008entanglement,casini2014remarks,aoki2015definition,ghosh2015entanglement,van2016entanglement,lin2020comments,chen2020strong}.%, e.g. algebraically or by embedding into a larger unphysical Hilbertspace (HS).
  \begin{figure}[t]
\begin{center}
\includegraphics[width=0.34\textwidth]{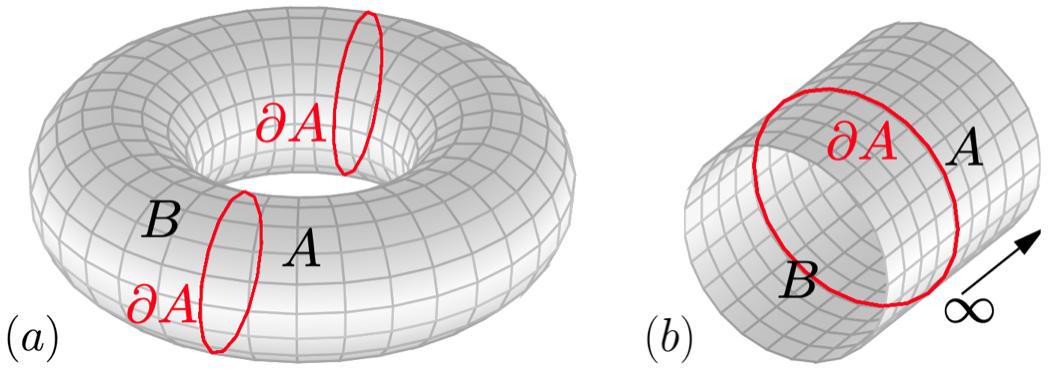}
\caption{Lattices with entanglement cuts considered in this work. (a) Torus and (b) infinite cylinder with entanglement boundary/boundaries $\partial A$.}
\label{fig:dualscheme}
\end{center}
\end{figure}

In this letter, we use dual theories~\cite{wegner1971duality,horn1979hamiltonian,radivcevic2016entanglement}  of  $\mathbf{Z}_2$ LGT in (2+1) spacetime dimensions ($\mathbf{Z}_2^{2+1}$)  embedded into a larger unphysical HS only along `entanglement' cuts, allowing access to the ES by naively `taking the trace'. 
%. This approach, in which naively `taking the trace' exactly reproduces what one expects algebraically~\cite{casini2014remarks,lin2020comments}, 
With this, we demonstrate Li and Haldane's en\-tangle\-ment-boundary conjecture for $\mathbf{Z}_2^{2+1}$ in the TO phase~\cite{li2008entanglement}, analytically on an infinite half-cylinder using perturbation theory and numerically on a finite torus at arbitrary coupling. A variational approach~\cite{kokail2020entanglement} allows us to reconstruct the Entanglement Hamiltonian (EH) of ground states, consistent with expectations from the Bisognano-Wichmann (BW) theorem~\cite{bisognano1975duality,bisognano1976duality}.

Our main effort is devoted to probing thermalization of LGTs through the ES. Focussing on out-of-equilibrium dynamics after quenches with initial states in the TO, as well as the trivial (confined) phase of the model,
we track the evolution of the symmetry resolved ES~\cite{rakovszky2019signatures,chang2019evolution,chamon2014emergent,yang2019scrambling,serbyn2016power,mierzejewski2013eigenvalue,bertini2019entanglement}: At early times the Schmidt rank is maximized, followed by spreading of level repulsion through the ES, and saturation of the entanglement entropy at parametrically later times. Remarkably, in an intermediate stage the approach to equilibrium is characterized by self-similarity of the ES, reminiscent of classical
wave turbulence and universal behavior~\cite{nazarenko2011wave,berges2014turbulent,berges2014basin,mace2020chiral}. %observed e.g. in the classical-statistical limit of Abelian and non-Abelian gauge theories far from equilibrium~\cite{berges2014turbulent,berges2014basin,mace2020chiral}.

%Thermalization differs between our two initial states, with the TO initial state, already fairly entangled, approaching (global) thermal equilibrium much quicker then the `confined' initial state.   

Despite being restricted to small systems, our ES analysis is remarkably robust  and provides a promising path towards understanding the thermalization of Abelian and non-Abelian gauge theories,  e.g. in Quantum Chromodynamics (QCD)~\cite{berges2020thermalization}. Our approach is suited for exploration with tensor networks~\cite{meurice2020tensor,banuls2020review,robaina2021simulating} in the case of ground and low energy states, as well as near-future  digital quantum computers and analog quantum simulators~\cite{kielpinski2002architecture,monroe2002quantum,blais2004cavity,cirac2012goals,hauke2012can,preskill2018quantum,martinez2016real,klco2018quantum,zache2018quantum,davoudi2020towards,mil2020scalable,de2021quantum} (see e.g.~\cite{schweizer2019floquet,barbiero2019coupling,homeier2021z} for  $\mathbf{Z}_2$ LGT).
 %It can be directly generalized to $\mathbf{Z}_n$ and $U(1)$ LGTs;  non-Abelian theories~\cite{douccot2004discrete,raychowdhury2020loop,davoudi2020search,ciavarella2021trailhead} will be explored in future work.  

%
%

%
\textit{Entanglement Structure of $Z_2$ Gauge Theory.}
We consider $\mathbf{Z}_2^{2+1}$  LGT with Hamiltonian
\begin{align}\label{eq:Z2Hamiltonian}
H= - \sum_{\mathbf{n}} \sigma^z_{\mathbf{n},x}\sigma^z_{\mathbf{n}+\hat{x},y}\sigma^{z}_{\mathbf{n}+\hat{y},x}\sigma^z_{\mathbf{n},y} -\epsilon\sum_{\mathbf{n},i=x,y} \sigma^x_{\mathbf{n},i} \,,
\end{align}
with $\mathbf{n}=(n_x,n_y)$, $n_i \in [0, N_i-1]$ and $i=x,y$
where $\sigma^z_{\mathbf{n},i}$ ($\sigma^x_{\mathbf{n},i}$) are Pauli operators positioned on the links of a two-dimensional rectangular lattice with $N_x\times N_y$ sites. Gauge invariance is expressed as $[H,G_\mathbf{n}]=0$ with 
$
G_\mathbf{n} = \sigma^x_{\mathbf{n},x}\sigma^x_{\mathbf{n}-\hat{x},x}
\sigma^x_{\mathbf{n},y}\sigma^x_{\mathbf{n}-\hat{x},y}\,
$
and Gauss law defines the physical subspace as $G_\mathbf{n} | \psi^{\rm phys} \rangle=| \psi^{\rm phys} \rangle$.

$\mathbf{Z}_2^{2+1}$  LGT has two ground state phases: a topologically trivial phase (confined), as well as a phase with topological order (TO). In the TO phase, for $\epsilon < \epsilon_c$, the ground state manifold is fourfold degenerate (on a torus), labelled by eigenvalues of `ribbon' operators $V_x \equiv \prod_{\mathbf{n}\in \mathcal{C}_x} \sigma^x_{\mathbf{n},i}$  and $V_y \equiv \prod_{\mathbf{n}\in \mathcal{C}_y} \sigma^x_{\mathbf{n},i}$, winding around the $x$- and $y$-directions, with $[V_x,H]=[V_y,H]=0$~\footnote{$V_x, V_y$ are electric flux loops winding around the periodic $x$- and $y$-directions of the torus, so called `ribbon' operators~\cite{sachdev2018topological,kitaev2003fault}, see also~\cite{michael1987glueball,van1987qcd}.}. 

In the following, we consider the entanglement properties of a bipartition $N_x\equiv N_x^A+N_x^B$ of a torus, see \Fig{fig:dualscheme}(a).
Before considering thermalization dynamics, we first validate our approach of computing LGT Entanglement Structure by demonstrating Li and Haldane's conjecture~\cite{li2008entanglement}:  TO 
is manifest in the entanglement structure of states;  the low lying part of the ES is equal (up to rescaling) to the physical spectrum of boundary excitations at the entanglement cut.

The basis of our analysis are dual formulations of \Eq{eq:Z2Hamiltonian} embedded into a larger, unphysical HS along boundaries: (a) for the torus with aforementioned entanglement bipartition [\Fig{fig:dualscheme}(a)], as well as (b)  an infinite (half) cylinder with physical boundaries [\Fig{fig:dualscheme}(b)], see Supplemental Material.  Our dual approach is a generalization of that of Wegner~\cite{wegner1971duality}, and unlike the latter where all Gauss laws are eliminated, it captures the entanglement structure stemming from the Gauss laws between the two subsystems, as we demonstrate below by verifying Li and Haldane's conjecture. By also resulting in a smaller Hilbert space, it reduces the numerical cost significantly compared to a direct implementation of \Eq{eq:Z2Hamiltonian}. 

To demonstrate Li and Haldane's conjecture, we consider first the semi-infinite cylinder A with physical `open electric' boundary conditions $\partial A$ in \Fig{fig:dualscheme}(b)~\cite{radivcevic2016entanglement,lin2020comments}. The corresponding dual Hamiltonian reads
 \begin{align}\label{eq:dualHamiltinfiniteTorus}
H_A=&- \sum_{\mathbf{n},n_x>0} \mu^z_{\mathbf{n}} - \sum_{\mathbf{n},n_x=0} \mu^z_{\mathbf{n}}\sigma^z_{\mathbf{n},y}
\nonumber\\
&-\epsilon \sum_{\mathbf{n},n_x>0}  \mu^x_{\mathbf{n} }\mu^x_{\mathbf{n}-\hat{x}} -\epsilon \sum_{\mathbf{n},n_x=0}  \sigma^x_{\mathbf{n},y}
\nonumber\\
&-\epsilon \sum_{\mathbf{n},n_y>0} \mu^x_{\mathbf{n} }\mu^x_{\mathbf{n}-\hat{y}}
-\epsilon \sum_{\mathbf{n},n_y=0} \mu^x_{\mathbf{n} }\mu^x_{\mathbf{n}-\hat{y}}V_y
\nonumber\\
&-\epsilon \sum_{n_y=1}^{N_y-1} \sigma^x_{{\scriptscriptstyle (-1,n_y)},x}\,.
\end{align}
This contains two sets of Pauli operators: dual gauge invariant operators $\mu^{x/z}_{\mathbf{n}}$  in the bulk as well as the original
 gauge-variant variables $\sigma^{x/z}_{\mathbf{n},i}$ on $\partial A$.  Here, $\sigma^x_{{\scriptscriptstyle (-1,n_y)},x}$ is the electric flux through the boundary, see also Supplemental Material. While gauge-redundancy is eliminated in the bulk, Gauss law on $\partial A$ is not  eliminated,  $G_{n_y}\equiv \sigma^x_{{\scriptscriptstyle (-1,n_y)},x} \sigma^x_{{\scriptscriptstyle (0,n_y)},y}  \sigma^x_{{\scriptscriptstyle (0,n_y-1)},y} \mu^x_{\mathbf{n} }\mu^x_{\mathbf{n}-\hat{y}} $ (for $n_y>0$) and  $\sigma^x_{{\scriptscriptstyle (-1,n_y)},x} \sigma^x_{{\scriptscriptstyle (0,n_y)},y}  \sigma^x_{{\scriptscriptstyle (0,n_y-1)},y} \mu^x_{\mathbf{n} }\mu^x_{\mathbf{n}-\hat{y}} V_y $ (for $n_y=0$).
 
To demonstrate Li and Haldane's conjecture, we first compute the boundary theory. The  ground state for $\epsilon =0$, $V_y=1$ is given by  $| \Omega^A \rangle = P_G | \ua \rangle $ where 
  $ | \ua \rangle$ is the state with plaquette eigenvalue $1$, i.e.
$ \mu^z_{\mathbf{n}}| \ua \rangle=| \ua \rangle$ in  the bulk ($n_x>0$) and  $ \mu^z_{\mathbf{n}}\sigma^z_{\mathbf{n},y}| \ua \rangle=| \ua \rangle$ on $\partial A$ ($n_x=0$); $P_G \equiv \prod_{n_y} (1+G_{n_y})/2$ is a projector onto the physical subspace. 
We compute the ground state for small $\epsilon$ perturbatively~\cite{bravyi2011schrieffer}
 (see \cite{patkos1981variational,dass1982phase,patkos1982improved,chen2020strong} for the opposite limit), resulting in a 
 low energy effective Hamiltonian $H_A^{\rm  eff}$ describing  excitations on $\partial A$, see Supplemental Material,
\begin{align}\label{eq:lowenergyH}
H_A^{\rm  eff} = -\epsilon \sum_{n_y =0}^{N_y-1}  
\sigma^x_{{\scriptscriptstyle (0,n_y)},y}  \sigma^x_{{\scriptscriptstyle (0,n_y-1)},y} \;\mu^x_{{\mathbf{n} }}\; \mu^x_{{\mathbf{n}-\hat{y}}} 
+ O(\epsilon^2)\,.
\end{align}
Here, we omitted the projector onto $| \ua \rangle$, given by $\prod_{\mathbf{n}} (1+W_\mathbf{n})/2$ with $W_\mathbf{n} = \mu^z_{\mathbf{n}} \sigma^z_{{\scriptscriptstyle (0,n_y)},y}$ ($n_x=0$) and  $W_\mathbf{n} =   \mu^z_{\mathbf{n}}$ ($n_x>0$), and constant terms.
\begin{figure}[t]
\begin{flushleft}
\qquad\qquad\includegraphics[width=0.28\textwidth]{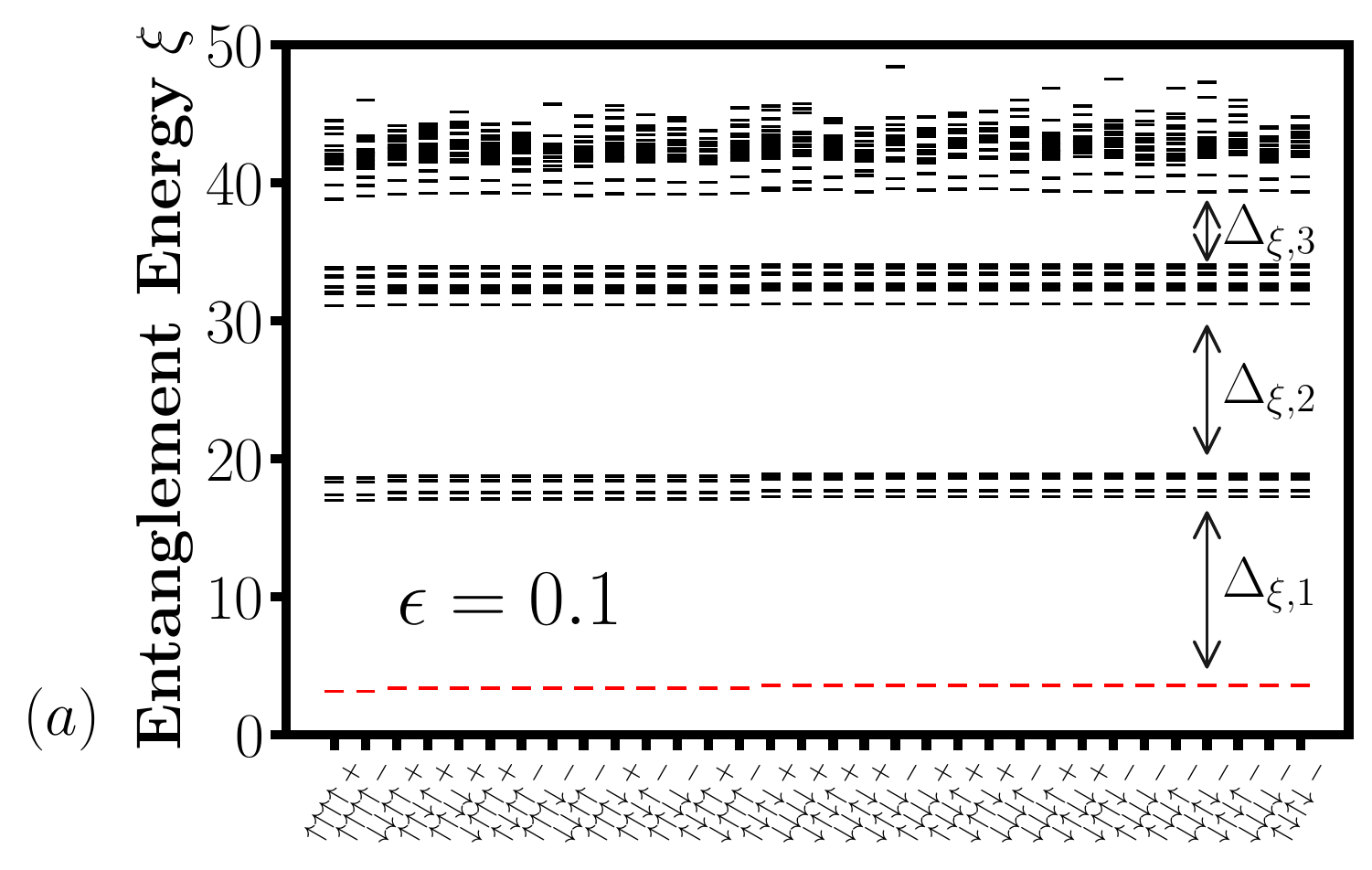}\\
\qquad\qquad\includegraphics[width=0.28\textwidth]{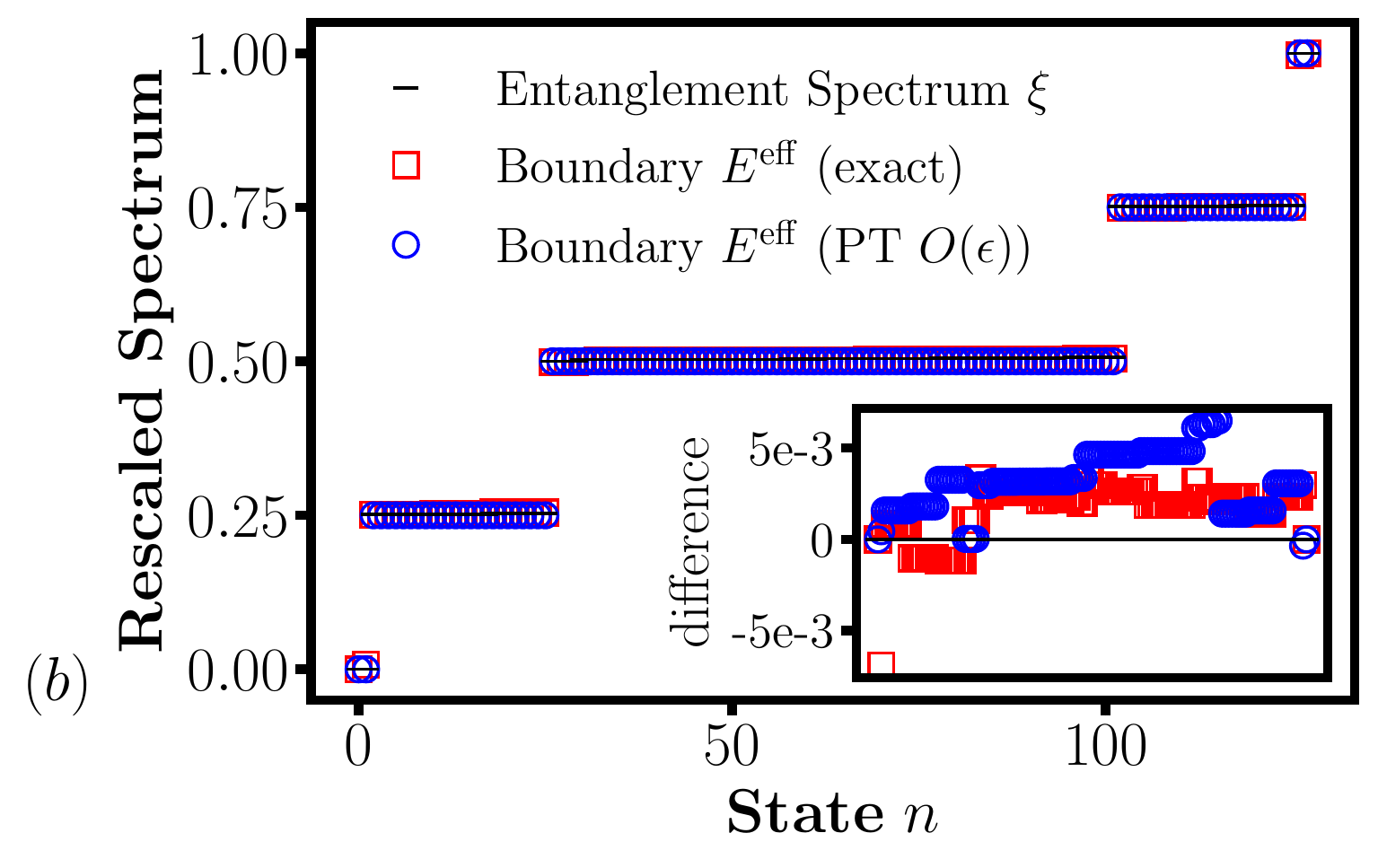}\\
\qquad\qquad\includegraphics[width=0.3\textwidth]{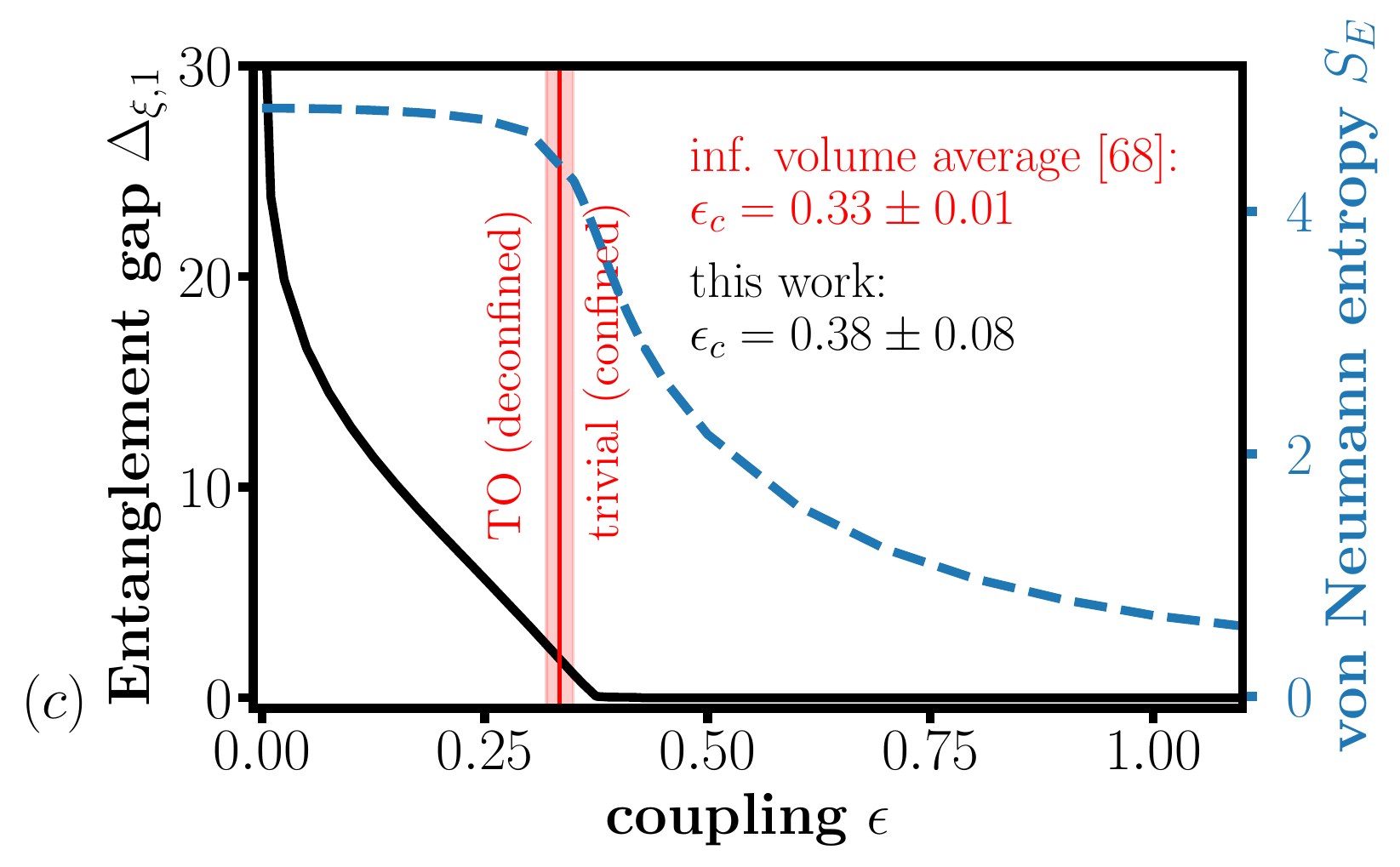}
\caption{(a) Entanglement Spectrum $\xi_n$ of the groundstate for $\epsilon=0.1$ and $(N_x^A + N_x^B) \times N_y = (3+3)\times 3$, corresponding to the entanglement cut in \Fig{fig:dualscheme}(a), resolved into symmetry sectors. (b) Rescaled low energy part of the entanglement spectrum for $\epsilon=0.1$, and $(N_x^A + N_x^B)\times N_y= (3+3)\times 4$, versus the spectrum of $H_A^{\rm eff}$: numerical (exact diagonalization)
and analytical (perturbation theory). (c) Entanglement gap $\Delta_{\xi,1}$ between low and high energy parts of ES (black), and von-Neumann Entropy (blue) as a function of coupling $\epsilon$ for $(N_x^A+N_x^B) \times N_y = (3+3)\times 4$ versus infinite volume average $\epsilon_c$ from  table II of \cite{blote2002cluster}.}
\label{fig:ES}
\end{flushleft}
\end{figure}

We now compute the ES for $\partial A$ a (virtual) entanglement cut of the infinite cylinder $A+B$, see~\Fig{fig:dualscheme}(b).
The density matrix of the ground state is [at $O(\epsilon)$]
\begin{align}\label{eq:rhoAB}
\rho_{{A+B}} =& \Bigg(\frac{1}{2} 
- \frac{\epsilon}{4} \Big[ \sum_{\substack{\mathbf{n},i \\  \not\in \partial A } } \mu^x_{{\mathbf{n}}} \mu^x_{{\mathbf{n}-\hat{i}}}  + \sum_{{n_y=0} }^{{N_y-1}} \sigma^x_{{\scriptscriptstyle(0,n_y)},\scriptscriptstyle{y}}  \Big] \Bigg)\rho^{(0)}
 +\rm{ h.c.}\,,
\end{align}
with $\rho^{(0)}\equiv \prod_{n_y} (1+G_{n_y})/2 \, \prod_{\mathbf{n}} (1+W_\mathbf{n})/2$,  where $W_\mathbf{n} = \mu^z_{\mathbf{n}} \sigma^z_{{\scriptscriptstyle (0,n_y)},y}$ for $n_x\in \{0,-1\}$ and  $W_\mathbf{n} = 
 \mu^z_{\mathbf{n}}$  for $n_x < -1$ and  $n_x>0$. In \Eq{eq:rhoAB}, ${}_{\mathbf{n},i\not\in \partial A}$ indicates summation over links in $A+B$ away from the entanglement cut. The reduced density matrix of system $A$ follows as
\begin{align}\label{eq:redDM}
&{\rho}_A 
= \Bigg (\frac{1}{2} -  \frac{\epsilon}{4} 
 \Big[ \sum_{\substack{\mathbf{n},i \in A\\  \not\in \partial A }} \mu^x_{\mathbf{n}} \mu^x_{\mathbf{n}-\hat{i}}  + \sum_{n_y=0}^{N_y-1} \sigma^x_{{\scriptscriptstyle(0,n_y)},y}  \,
 \nonumber\\
& -\sum_{n_y=0}^{N_y-1} \mu^x_{{\scriptscriptstyle(0,n_y)}}\mu^x_{{\scriptscriptstyle(0,n_y-1)}}
 \sigma^x_{{\scriptscriptstyle(0,n_y)},y}\sigma^x_{{\scriptscriptstyle(0,n_y-1)},y} \Big]\,  \Bigg) \rho^{(0)}_A +{ \rm h.c.},
\end{align}
with $\rho^{(0)}_A\equiv  \mathbb{I}^{\partial A} /2^{N_y} \prod_{\mathbf{n} \in A} (1+W_\mathbf{n})/2$ and $\mathbb{I}^{\partial A} $
a $2^{N_y}$ dimensional unit matrix on  $\partial A$.  We thus obtain  the EH $H_A^{\rm ent}=-\log(\rho_A)$   from \Eq{eq:redDM},
\begin{align}\label{eq:EHpt}
&H_A^{\rm ent}
= N_y \log(2) + \frac{\epsilon}{2} \sum_{n_y =0}^{N_y-1}  
\sigma^x_{{\scriptscriptstyle (0,n_y)},y}  \sigma^x_{{\scriptscriptstyle (0,n_y-1)},y} \;\mu^x_{{\mathbf{n}} }\; \mu^x_{{\mathbf{n}-\hat{y}}} \,,
\end{align}
omitting again the projector $ \prod_{\mathbf{n} \in A} (1+W_\mathbf{n})/2$.
\Eq{eq:EHpt} has precisely the same form as \Eq{eq:lowenergyH}, $H_A^{\rm ent}\simeq H_A^{\rm eff}$,  demonstrating (perturbatively) Li and Haldane's conjecture for $\mathbf{Z}^{2+1}_2$ lattice gauge theory.% (up to rescaling and constants determined by normalization). 
\begin{figure}[t]
\begin{center}
\includegraphics[width=0.235\textwidth]{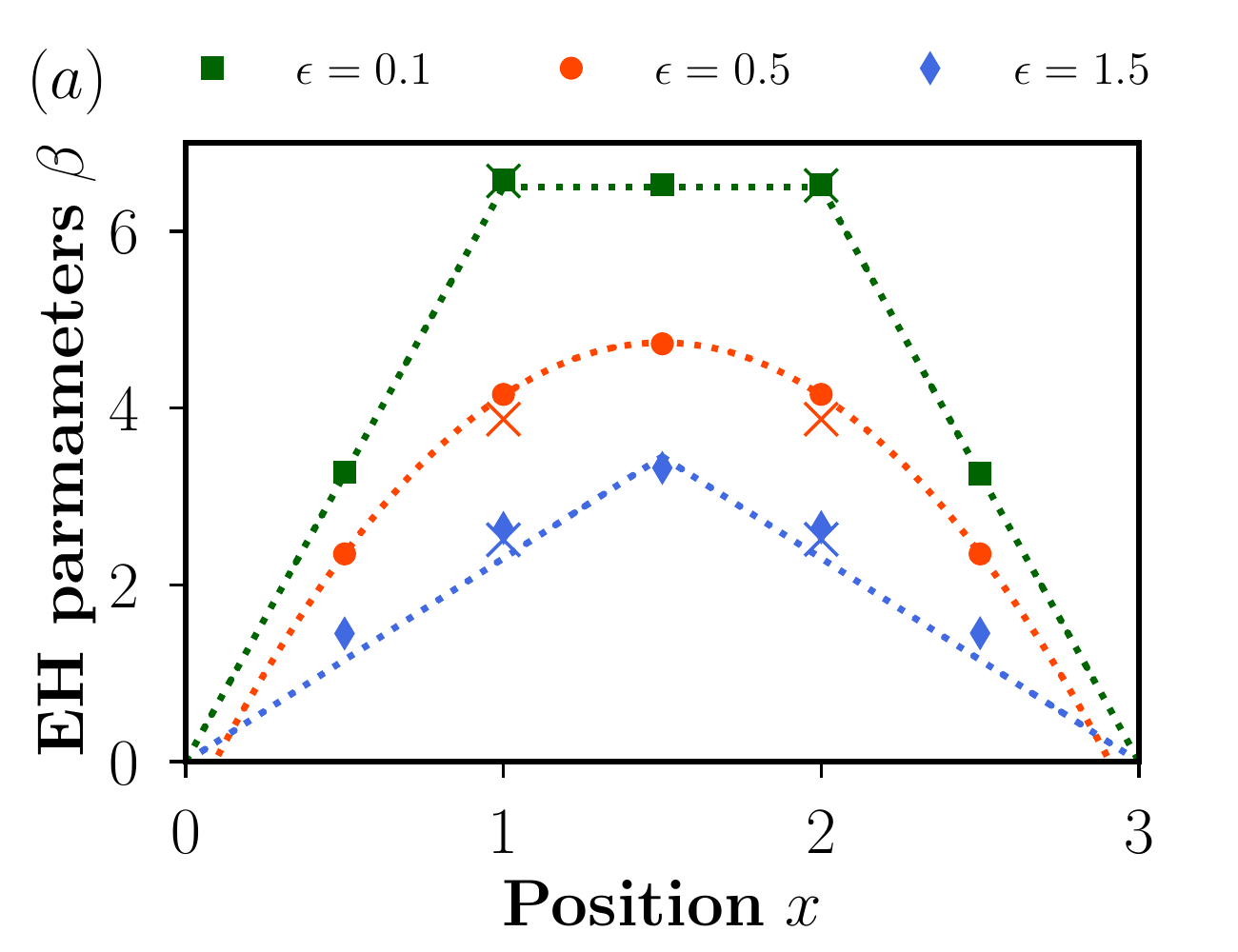}
\includegraphics[width=0.235\textwidth]{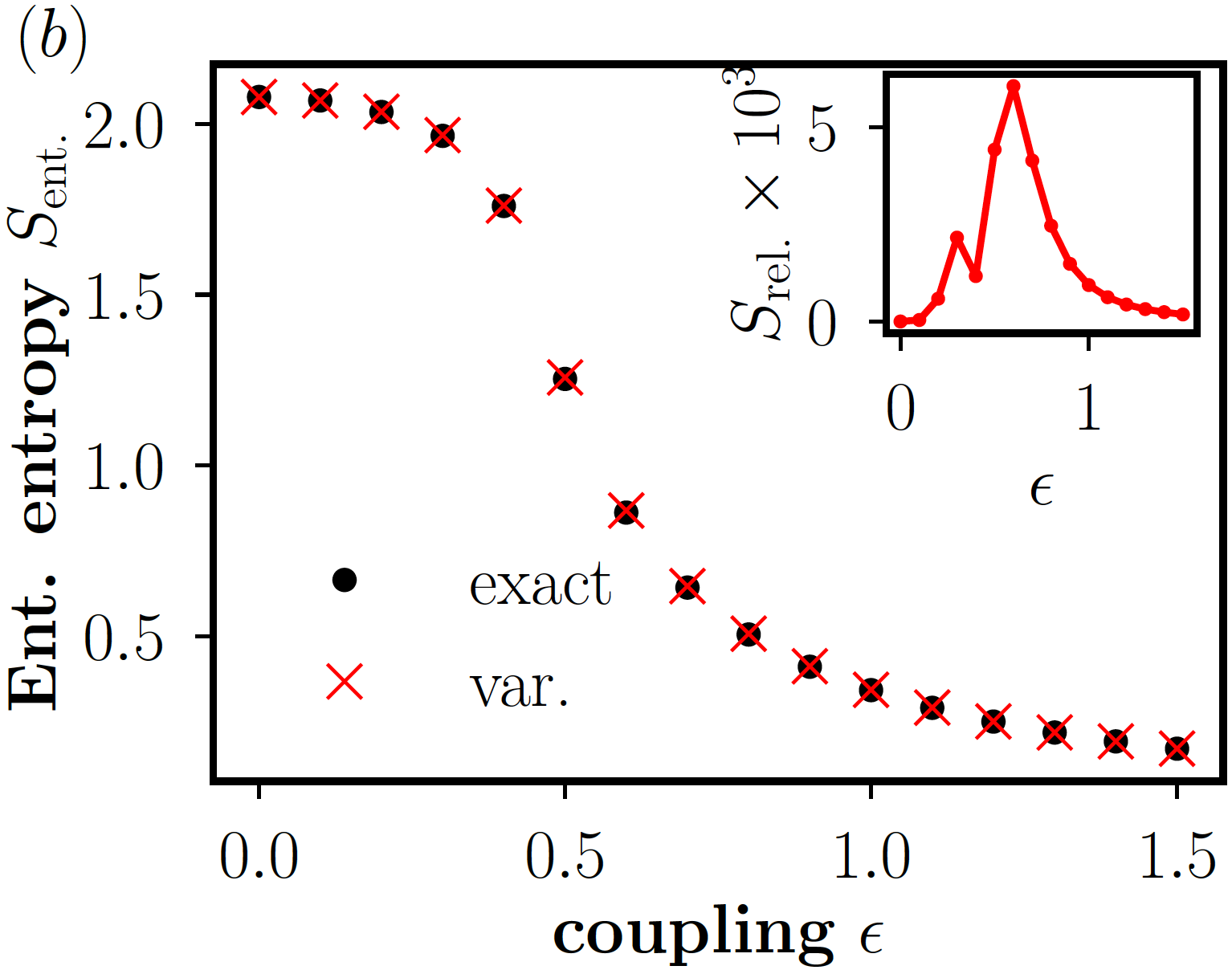}
\includegraphics[width=0.47\textwidth]{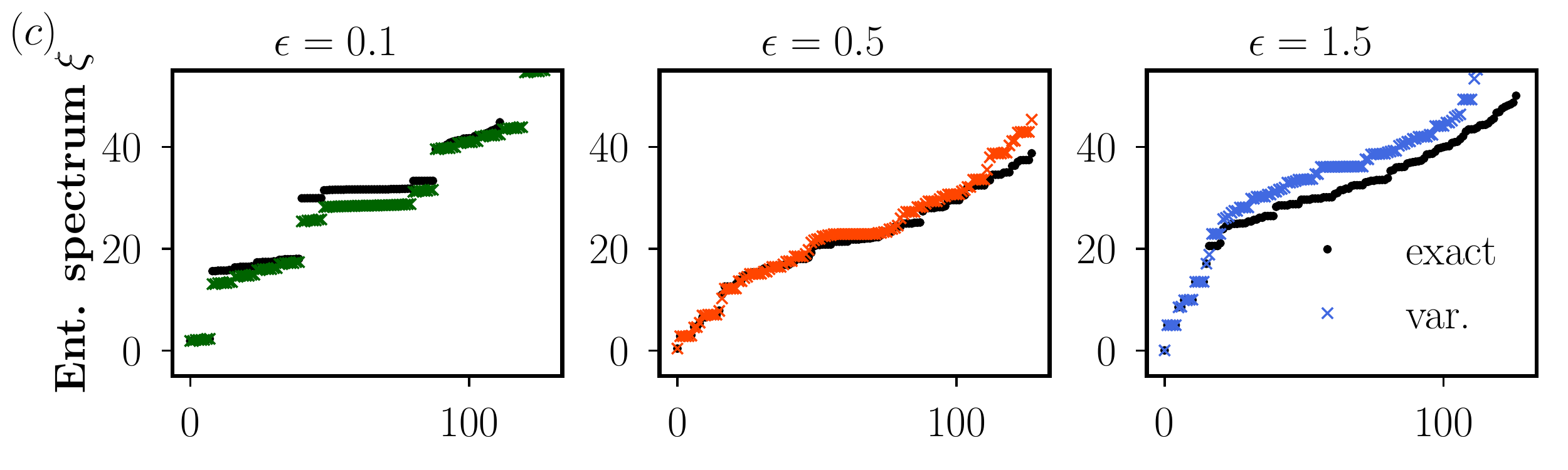}
\caption{(a) Optimal EH parameters for the local approximation $\sigma_A$. Thick markers and light crosses correspond to electric and magnetic energy contributions, respectively. The dotted, orange line is a parabolic fit; the other dotted, green and blue lines are guides for the eye. (b) Exact (black dots) and optimal variational (red crosses) entanglement entropy. Inset: relative entropy. $(c)$ Exact (black dots) and optimal variational (colored crosses) ES for the same values of $\epsilon$ shown in (a). All data are obtained for a (sub)system of size $(N_x^A+N_x^B)\times N_y = (3+3)\times 2$.}
\label{fig:BW}
\end{center}
\end{figure}

To probe the validity of this equivalence beyond perturbation theory, we turn to numerical simulations with exact diagonalization~\cite{weinberg2018quspin}. We consider a torus separated by two entanglement cuts into systems $A+B$, shown in \Fig{fig:dualscheme}(a).  The ES $\xi_n$ of states $n=1,\dots,{\rm{\dim }}(\rho_A)$ in $A$ is shown in
\Fig{fig:ES}(a), separated into symmetry sectors, specified by the electric flux operators into the system on both boundaries ($\ua/\da$) and a string of electric field operators \ $\tilde{V}_x = \prod_{\mathbf{n}\in \tilde{\mathcal{C}} } \sigma^x_{\mathbf{n},i}$ across a path $\tilde{\mathcal{C}}$ from one boundary to the other ($\pm$), see Supplemental Material. \Fig{fig:ES}(b) shows the smallest eigenvalues (`low energy part') of the ES compared to the boundary spectrum of $H_A$  for $\epsilon=0.1$, displaying near perfect agreement with each other and with our perturbative result. This agreement holds in the TO phase with good precision up to finite size corrections. 

Another signature of TO are Entanglement Gaps of the ES $\Delta_{ \xi,{i}}$ shown in \Fig{fig:ES}(a) and (c), the latter displaying $\Delta_{ \xi,{1}}$  as a function of $\epsilon$.
Defining the TO/confinement phase transition $\epsilon_c$ at  $\Delta_{ \xi,{1}}\rightarrow 0$ results in $\epsilon_c = 0.38 \pm 0.09$ agreeing within errorbars with the infinite volume result $ \epsilon_c = 0.33 \pm 0.01$~\cite{blote2002cluster}, see Supplemental Material where we demonstrate robustness against finite-volume effects.

The Bisognano-Wichmann (BW) theorem, and its extensions~\cite{bisognano1975duality,bisognano1976duality}, captures another aspect of the EH; it states that the EH of the ground state is a local deformation of the system Hamiltonian. 
Using an ansatz to approximate the reduced density matrix $\rho_A \approx \sigma_A \propto \exp \left(-\sum_{\mathbf{n}\in A} \beta_\mathbf{n}h_\mathbf{n}\right)$, with $h_\mathbf{n}$ denoting gauge-invariant local operators in $A$, we test its applicability to LGTs. Optimal local parameters $\beta_\mathbf{n}$, obtained by minimizing the relative entropy $S(\rho_A|| \sigma_A)$, are shown in~\Fig{fig:BW}(a), see also Supplemental Material.

\begin{figure}[t]
\begin{center}
\quad\;\; \includegraphics[width=0.32\textwidth]{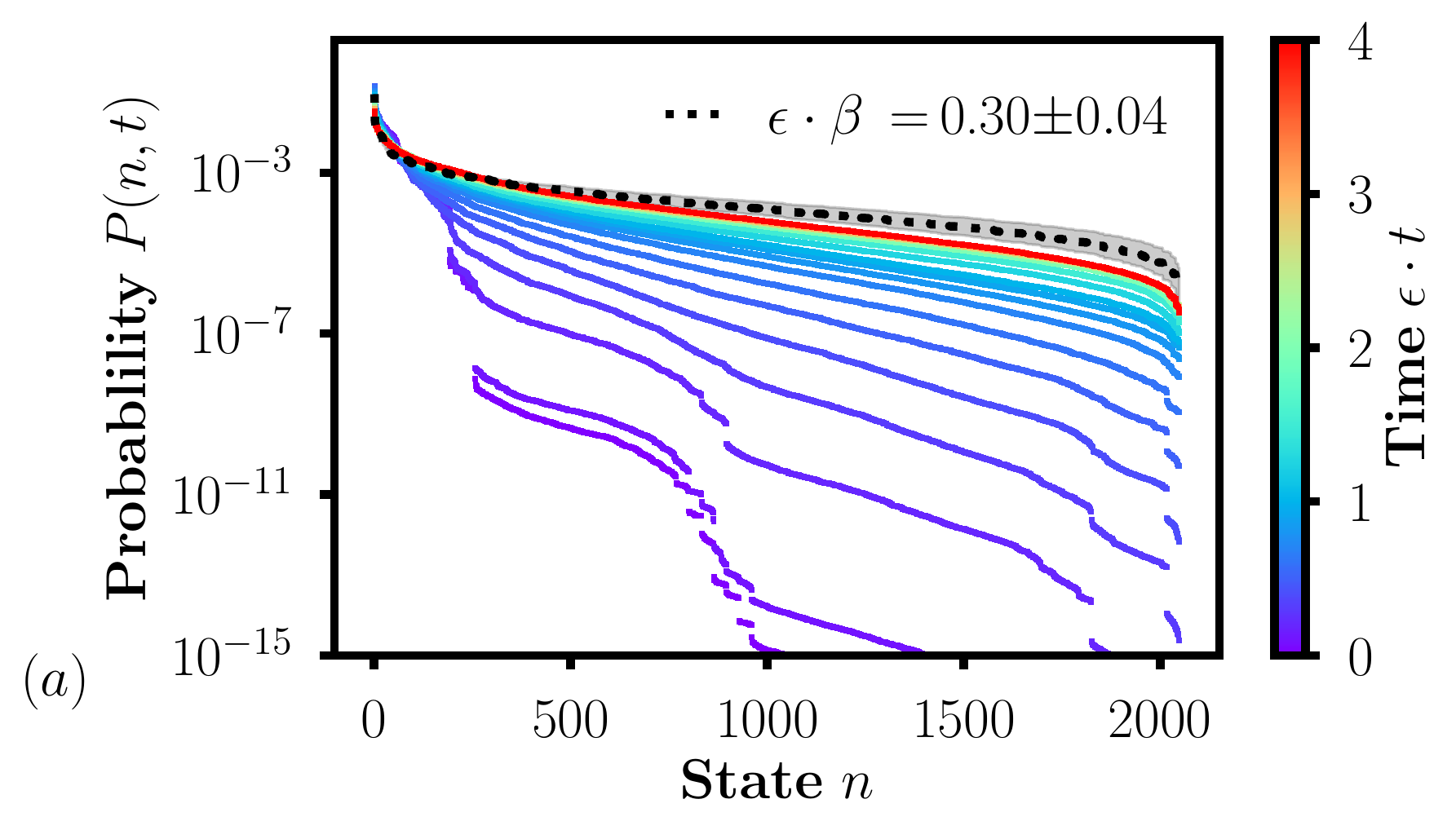}\qquad\quad{}\\
\;\quad \includegraphics[width=0.31\textwidth]{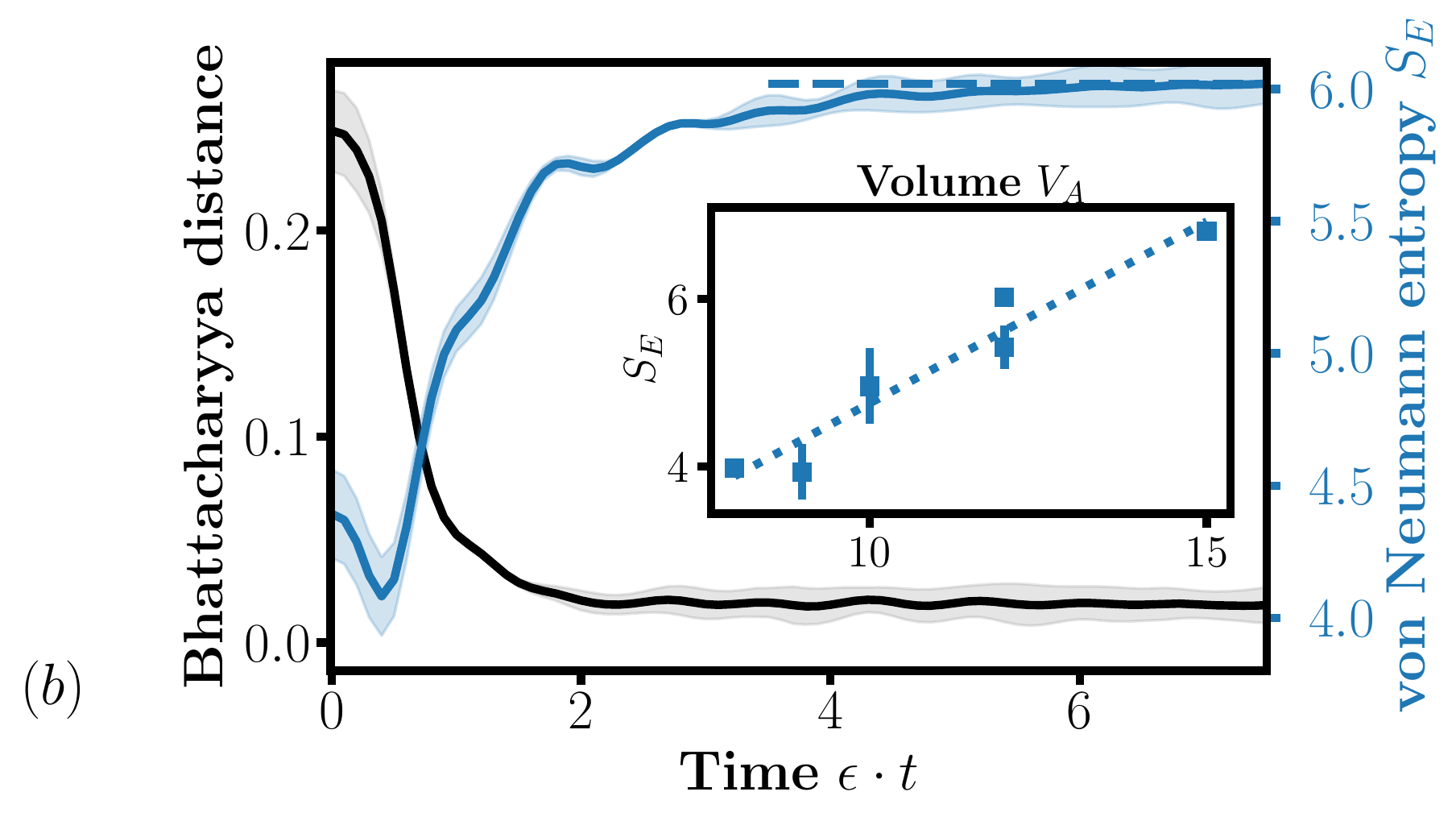}\\
\includegraphics[width=0.29\textwidth]{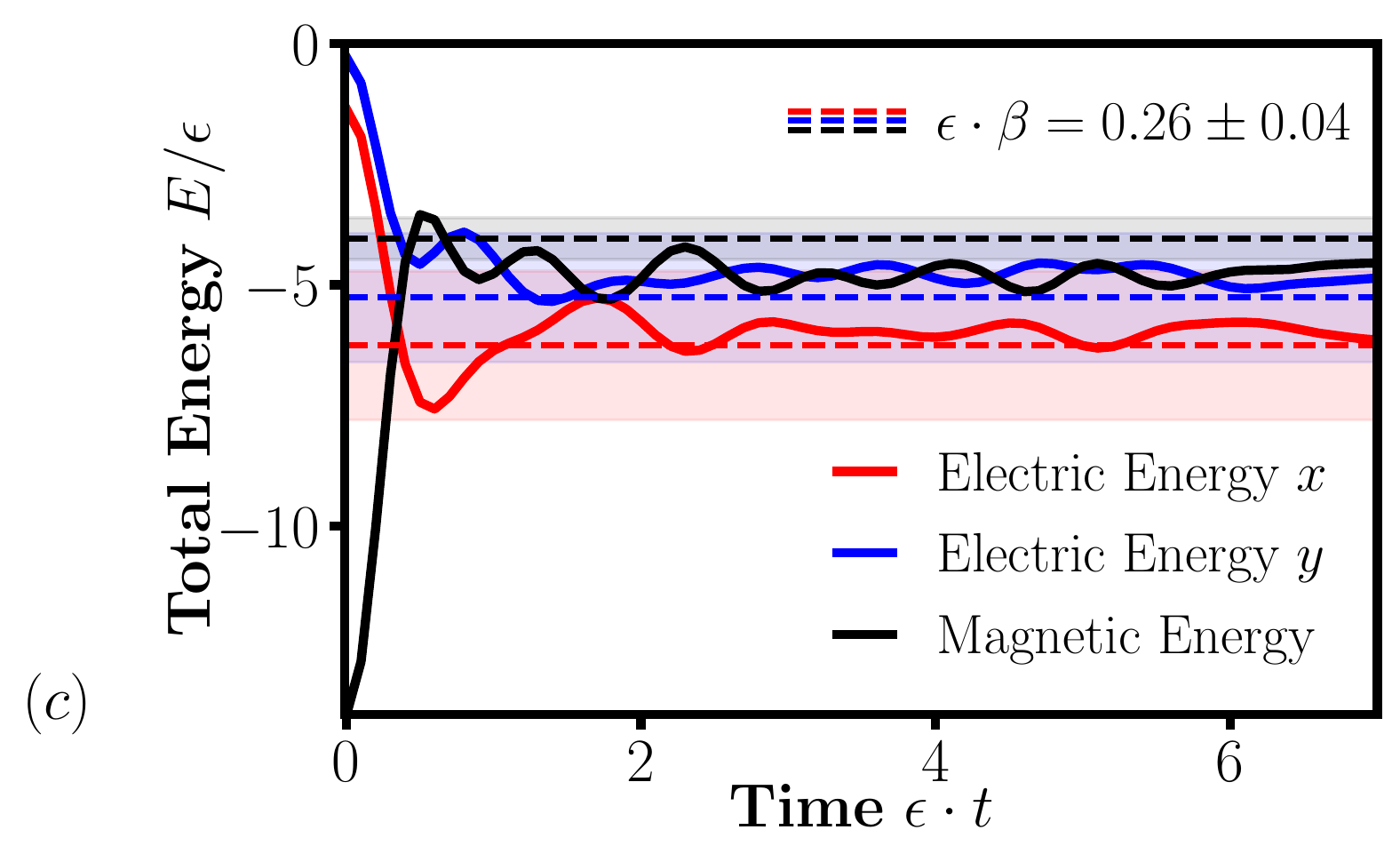}
\caption{(a) ES as a function of time ($(N_x^A+N_x^B) \times N_y=(3+5)\times 3$) versus that of a thermal system (black dotted line). (b) Von Neumann entropy and Bhattacharyya distance to a thermal system. Inset: Dependence of the saturated entropy on the volume $V_A$ of $A$. (c) Electric and magnetic energy compared to thermal expectation values ($N_x\times Ny = (3+3)\times 3$). In (a) $\beta$ is determined from the saturated entanglement entropy in (b), while in (c) it is determined from the energy density of the initial state. Bands and error bars indicate uncertainties due to  finite-size effects, determined from the difference between largest and second-largest lattices.}
\label{fig:thermalization}
\end{center}
\end{figure}

\begin{figure*}[t]
\centering{
\includegraphics[scale=0.38]{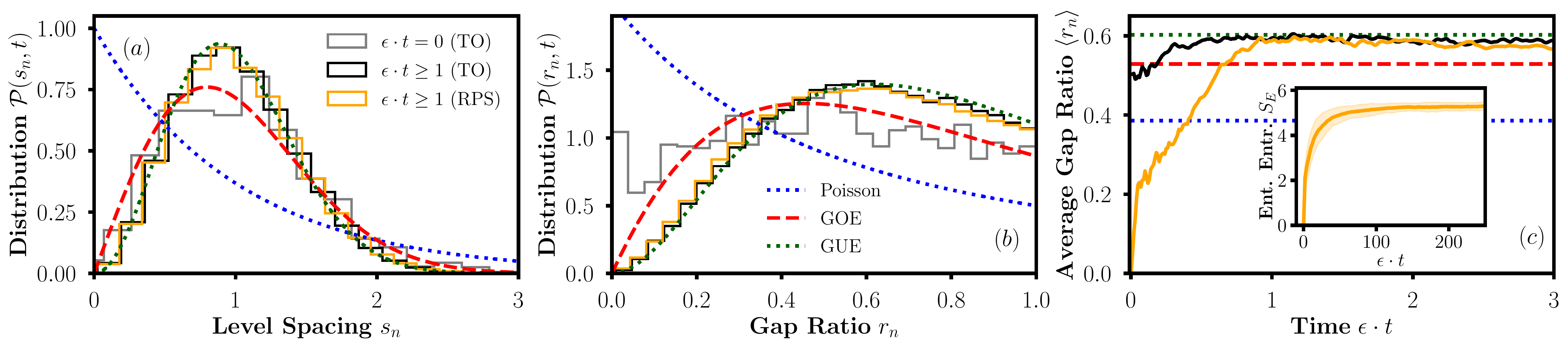}
}
\caption{(a) Distribution $\mathcal{P}(s_n)$ of level spacings (of the unfolded ES), for the quench from $\epsilon =0.1 \rightarrow 1$ (gray and black curves) versus $\epsilon \rightarrow \infty$ (confined phase) as initial condition (orange curve). (b) Gap ratio distribution $\mathcal{P}(r_n)$. (c) Average gap ratio $\langle r_n \rangle$ as a function of time. Inset: Growth of von Neumann entropy for the quench $\epsilon = \infty  \rightarrow 1$. (Shown for $(N_x^A+N_x^B)\times(3+5)\times 3 $ lattice sites.)}\label{fig:levelspacing}
\end{figure*}

Our results are consistent with a parabolic deformation in the vicinity of the phase transition, as expected from BW for a conformal field theory.
While the deformation deviates from a parabola away from the critical point, the overall quality of the approximation $H^\text{ent}_A \approx - \log \sigma_A$ is excellent, see Fig.~\ref{fig:BW}(b) and Fig.~\ref{fig:BW}(c), comparing the exact ES to its variational approximation. The local deformation captures the low energy part of the ES almost perfectly for all $\epsilon$, except for small deviations at high energy which do not contribute significantly to the entanglement entropy. %In the vicinity of the critical point, we find excellent agreement also in the ``high-energy'' part of the ES, indicating that the lattice BW theorem becomes more accurate or even exact in this case.
 \begin{figure}[t]
\begin{center}
\includegraphics[width=0.45\textwidth]{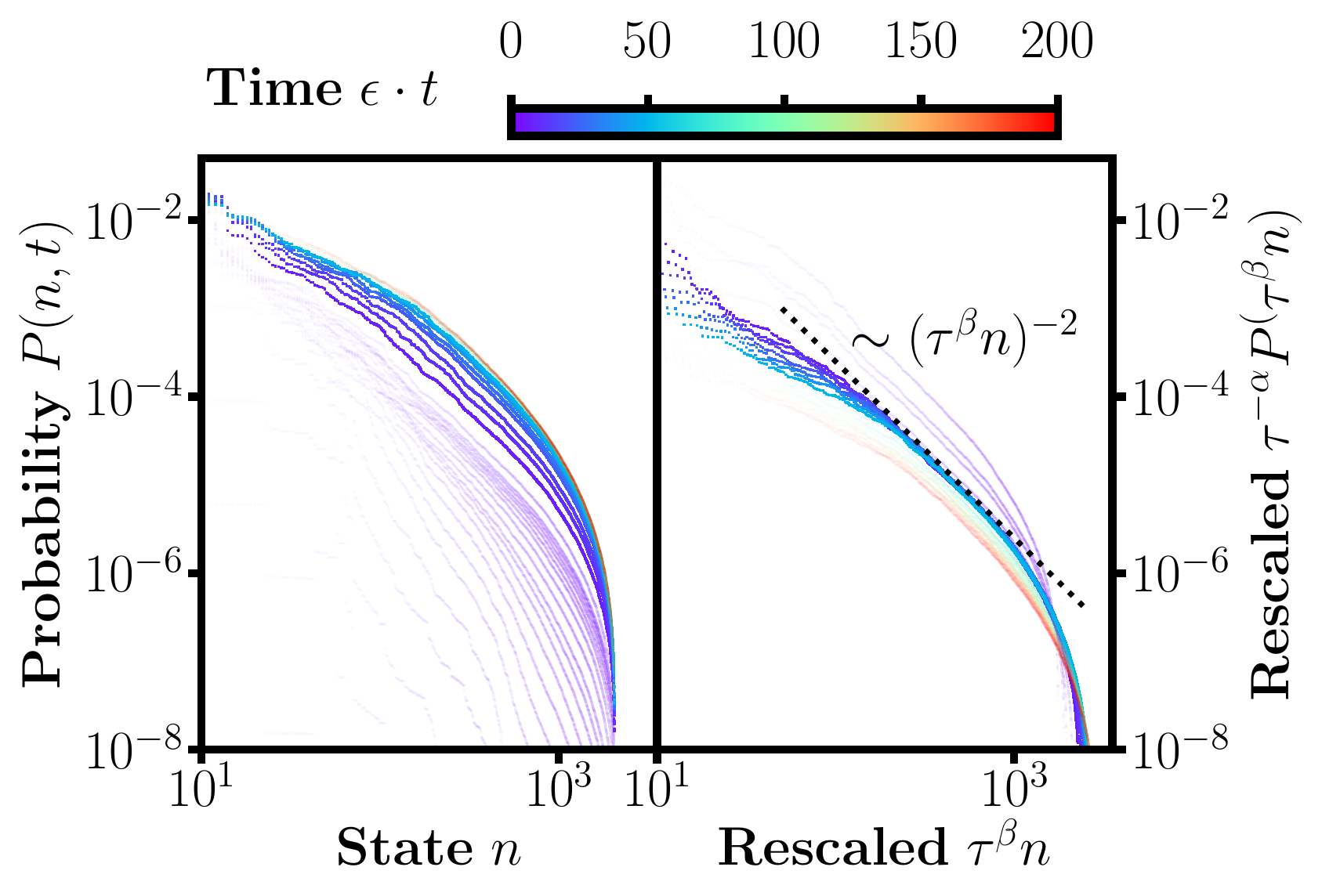}
\caption{Left: Un-rescaled Schmidt spectrum $P(n,t)=\exp \{ -\xi_n\}$ for the quench $\epsilon = \infty \rightarrow 1$  at different times. Right: Rescaled Spectrum. The approach to thermalization is characterized by a self-similar universal form $P(n,t) = \tau^{-\alpha} P( \tau^\beta n )$, $\tau \equiv \epsilon (t-t_0)$ for times $ 2 \lessapprox \epsilon \cdot t \lessapprox 60-100$. A black dotted line indicates power law behavior $(\tau^\beta n )^{-2 }$. The spectrum outside the scaling window is shaded out.  (Shown for $(N_x^A+N_x^B)\times(3+5)\times 3 $ lattice sites.)}
\label{fig:universal}
\end{center}
\end{figure}

\textit{Thermalization from the Entanglement Spectrum.} 
To characterize thermalization dynamics, we extract the ES of non-equilibrium states in the following. We prepare a (randomly chosen) excited eigenstate of $H(\epsilon=0.1)$ as initial state and evolve with $H(\epsilon =1)$. \Fig{fig:thermalization}(a) demonstrates thermalization, showing the corresponding Schmidt spectrum $P(n,t)\equiv\exp\{ - \xi_n(t)\}$ of subsystem $A$  [$N_x \times N_y =  (3+5)\times 3$] approaching the thermal limit (black dotted line) at late times. %This initial state is already fairly entangled, see the case of an initial product state below. 
Thermalization occurs when expectation values are equal to those derived from a canonical ensemble, i.e.  $\rho^{\rm therm.}_A = \text{Tr}_B (\rho^{\rm can.} ) / \text{Tr} ( \rho^{\rm can.})$ where $\rho^{\rm can.}=\exp\{ -\beta H\}$~\cite{garrison2018does}.
In \Fig{fig:thermalization}(a), we use an approximate, but numerically simpler, form $\rho_A^{\rm therm.} \approx e^{- \beta H_A} /   \text{Tr}_A ( e^{-\beta H_A} )$~\cite{garrison2018does}, with $H_A$  the projection of the Hamiltonian onto subsystem $A$  [boundaries are  as in \Eq{eq:dualHamiltinfiniteTorus}]. In \Fig{fig:thermalization}(b), we show the  Bhattacharyya distance~\cite{bhattacharyya1943measure} between $\rho_A$ and $\rho_A^{\rm therm.}$ and the von Neumann entropy, whose saturated value exhibits a volume law as displayed in the inset.

The dynamics of electric and magnetic energies is shown in \Fig{fig:thermalization}(c) and compared with their thermal expectations, for system size $N_x \times N_y=(3+3)\times 3$. Here, $\rho_A^{\rm therm.}$ is determined from the exact $\rho^{\rm can.}$. To estimate the systematic error resulting from the approximate form of $\rho_A^{\rm therm.}$ and finite volume effects, in (c) the inverse temperature $\beta\approx 0.26$ is determined by the total energy, while in (a) and (b) $\beta \approx 0.3$ it is determined from matching
the saturated entanglement entropy to the thermal entropy of the same system $H_A$. %We note that the thermalization times $\epsilon \cdot t^{\rm therm.} \approx 3-4$ are consistent.

The ES allows to characterize the stages of the thermalization process. To show this, we consider the distribution $\mathcal{P}(s_n,t)$ of level spacings $s_n = \tilde{\xi}_n - \tilde{\xi}_{n-1}$ of the unfolded ES $\tilde{\xi}_n$~\cite{guhr1998random}, again resolved into symmetry sectors. Additionally, we consider the gap ratio~\cite{oganesyan2007localization}
\begin{align}\label{eq:gapratio}
r_n \equiv \frac{\text{min} (\delta_n,\delta_{n-1})}{\text{max} (\delta_n,\delta_{n-1})}
\end{align}   
where $\delta_n =\xi_n - \xi_{n-1}\ge 0$ of the ES $\xi_n$.
\Fig{fig:levelspacing}(a) shows the level spacing distribution $\mathcal{P}(s_n,t)$ at $t=0$ (gray) and for  $\epsilon \cdot t \ge 1$ (black), combined for all symmetry sectors. We compare this with  a completely uncorrelated Poisson distribution (blue dotted),
a Gaussian Orthogonal Ensemble (GOE, red dashed) and a Gaussian Unitary Ensemble (GUE, green dotted). Along with the distribution of the gap ratio $\mathcal{P}(r_n ,t)$ in \Fig{fig:levelspacing}(b), level statistics consistent with GUE is observed for $\epsilon \cdot t\gtrapprox 0.5$ in \Fig{fig:levelspacing}(c) (black curve), well before the thermalization time scale $\epsilon\cdot t^{\rm therm.}\approx 4$ seen in \Fig{fig:thermalization}(b).
 
In order to probe the independence of thermalization on the special initial state, with large entanglement and ES level repulsion,
we now consider a different scenario starting from an entirely unentangled initial state: a randomly chosen excited state of the $\epsilon \rightarrow \infty$ `electric ground state' (the confined phase). 
Orange curves
in \Fig{fig:levelspacing}(a-c) show the resulting level spacing and gap ratio approaching GUE at $\epsilon \cdot t\gtrapprox 1$ starting from the trivial ES at $\epsilon \cdot t =0$. The average gap ratio in (c) quickly jumps to $\langle r_n \rangle\approx 0.2$ at earliest times, then grows linearly until it saturates to about $\langle r_n \rangle \approx 0.6$. The inset of (c) shows that the growth of entanglement is much slower and saturates at parametrically later times $\epsilon \cdot t_{\rm therm.} \lessapprox 150$, similar to the separation observed in~\cite{rakovszky2019signatures} (see also \cite{calabrese2009entanglement}).

Remarkably, we find that the stage between the build-up of level repulsion and entanglement saturation is characterized by a self-similar scaling form of the spectrum $P(n,t)$, shown in \Fig{fig:universal}, reminiscent of classical wave turbulence~\cite{nazarenko2011wave,berges2014turbulent,berges2014basin,mace2020chiral}. In this regime, the spectrum can be rescaled as $P(n,t) = \tau^{-\alpha} P( \tau^\beta n)$ with $\tau \equiv \epsilon (t-t_0)$. We numerically determined $\epsilon \cdot t_0 = 1.8 \pm 0.5$ 
and the scaling coefficients, see Supplemental Material for details,
$$
\alpha = 0.8 \pm 0.2\,  \quad \beta = 0.0 \pm 0.1\,.
$$%
Our observation implies that thermalization occurs through turbulent transport of probability from the
`high energy' (small probability) towards the low lying part of the ES (large probability). The errors quoted for $\alpha,\beta$ include finite-volume and errors from the statistical procedure of extracting them; a detailed analysis, including results on larger lattices, can be found in Supplemental Material. 

% Thermalization is `quantum chaotic' in the sense of this `turbulent transport', of probability from the
% `high energy' (small probability) towards the low lying part of the ES (large probability).
% %
% %seems to proceed via `turbulent' transport of a conserved quantity (in this case probability) from the .
% %\NM{Is this the time when entanglement has spread through the entire subsystem and chaos happens on scales %larger than the subsystem, hence universal. See~\cite{rakovszky2019signatures}}
% This observation strongly hints at reconciliation of the (naively different) quantum versus classical  thermalization paradigms, i.e. in terms of  matrix elements of observables~\cite{deutsch1991quantum,srednicki1994chaos} versus
% ergodicty, chaos and universality~\cite{nazarenko2011wave}. Because time evolution in quantum mechanics is linear, quantum chaos is hidden in the complexities of energy eigenfunctions~\cite{srednicki1994chaos}, however, (and perhaps not so surprisingly~\cite{garrison2018does}) it is evident in the Entanglement Spectrum.

% \NM{Comments on interpretation a la kinetic theory, universality etc. Discussion of ETH and chaos and universality.}
%
%

\textit{Summary and Conclusions.} In this letter, we explored the entanglement structure of LGTs to characterize ground states, quantum phase transitions and thermalization, using dual theories of $\mathbf{Z}^{2+1}_2$ embedded into a larger gauge-variant HS only along entanglement boundaries~\cite{buividovich2008entanglement,casini2014remarks,aoki2015definition,ghosh2015entanglement,lin2020comments,chen2020strong}. Our fairly simple approach, see Supplemental Material for details, can be generalized to $\mathbf{Z}_n$ and $U(1)$ LGTs; non-Abelian theories~\cite{douccot2004discrete,raychowdhury2020loop,davoudi2020search,ciavarella2021trailhead,Wiese2021} are more challenging. Ising-like dualities~\cite{drouffe1979lattice,douccot2004discrete,mathur2015canonical}, prepotential-~\cite{mathur2005harmonic,mathur2007loop,anishetty20142,mathur2010n} and `Loop-String-Hadron'~\cite{raychowdhury2020solving} formulations are promising approaches, and will be explored in future work.

We demonstrated Li and Haldane's entanglement-boundary conjecture~\cite{li2008entanglement} for $\mathbf{Z}_2^{2+1}$ gauge theory, both analytically (in perturbation theory) and numerically using exact diagonalization. Moreover, we reconstructed the Entanglement Hamiltonians of ground states, finding consistency with expectations from the Bisognano-Wichmann theorem~\cite{kokail2020entanglement,bisognano1975duality,bisognano1976duality} at arbitrary coupling. 
Using the closing of the Entanglement Gap of the ES, we determine the confinement/deconfinement phase transition at  $\epsilon_c = 0.38\pm 0.09$. We find agreement within error bars with the infinite volume results, demonstrating the potential usefulness of Entanglement Structure, compared to computing volume versus boundary law scaling of Wilson loop operators.

Our most important result is that $\mathbf{Z}_2^{2+1}$ thermalization occurs in
 clearly separated stages: Starting from an initial (unentangled) product state, the system maximizes its Schmidt rank quickly, followed by rapid spreading of level repulsion
throughout the ES at early times. An intermediate regime is characterized by self-similar scaling of the Schmidt spectrum,  reminiscent of wave turbulence and universality in (semi-)classical systems, with scaling coefficients $\alpha = 0.8 \pm 0.2$, $\beta = 0.0 \pm 0.1$.

This observation strongly hints at a reconciliation of the (naively different) quantum versus classical  thermalization paradigms, i.e. in terms of  matrix elements of observables~\cite{deutsch1991quantum,srednicki1994chaos} versus
ergodicity, chaos and universality~\cite{nazarenko2011wave}. Because time evolution in quantum mechanics is linear, quantum chaos is hidden in the complexities of energy eigenfunctions~\cite{srednicki1994chaos}, however, (and perhaps not so surprisingly~\cite{garrison2018does}) it becomes evident in the Entanglement Spectrum.
Our analysis provides a systematic path for the quantification and classification of this behavior, which is likely generic for gauge and non-gauge systems and in line with the ETH. Our numerical investigations are not exhaustive, and could be extended to, e.g., studying the build-up of volume law entanglement, spectral form factors~\cite{chang2019evolution,guhr1998random}, or higher order level spacing ratios~\cite{tekur2020symmetry} of the ES.  It would also be interesting to apply our techniques to systems with many-body localization~\cite{brenes2018many}.

Apart from the importance of (2+1)d LGTs for, e.g., topological quantum computation~\cite{kitaev2003fault,satzinger2021realizing}, and condensed matter physics~\cite{kasper2020jaynes,wen2004quantum}, the Entanglement structure of Abelian and non-Abelian gauge theories, such as QCD, may be crucial for thermalization in high energy and nuclear physics, where it is largely unexplored. Examples are the apparent quick thermalization and hydrodynamization of the Quark-Gluon-Plasma in ultra-relativistic heavy ion collisions~\cite{berges2020thermalization} or the structure of QCD bound states in deeply inelastic scattering experiments~\cite{kharzeev2017deep,mueller2020deeply,tu2020einstein}. %In future work we will consider extensions  to non-Abelian gauge theories.
%\NM{Approach to QGP, quick thermalization}
%Doing so, for more complicated theories and larger systems, may be impossible with classical lattice techniques in Euclidean spacetime, but should be explored with tensor network techniques~\cite{meurice2020tensor,banuls2020review,robaina2021simulating}. 

Understanding thermalization of quantum many-body systems, in particular of gauge theories, is a unique opportunity for quantum computers and analog 
simulators~\cite{kielpinski2002architecture,monroe2002quantum,blais2004cavity,cirac2012goals,hauke2012can,preskill2018quantum,martinez2016real,klco2018quantum,zache2018quantum,davoudi2020towards,mil2020scalable,de2021quantum,schweizer2019floquet,barbiero2019coupling,homeier2021z}.
Entanglement structure of quantum many-body states can be extracted in state-of-the-art quantum simulation experiments, see e.g. \cite{pichler2016measurement,dalmonte2018quantum,elben2020mixed,kokail2020entanglement}.

\textit{Acknowledgements.} N.M. thanks Zohreh Davoudi and Nikhil Karthik for valuable discussions. 
N.M. acknowledges funding by the U.S. Department of Energy’s Office of Science, Office of Advanced Scientific Computing Research, Accelerated Research in Quantum Computing program award DE-SC0020312. N.M. also acknowledges support by the U.S. Department of Energy, Office of Science, Office of Nuclear Physics, under contract No. DE-SC0012704
and by the Deutsche Forschungsgemeinschaft (DFG, German Research Foundation) - Project 404640738 during early stages of this work. R.O. thanks J\"urgen Berges for fruitful discussions.
R.O. acknowledges funding from the DFG (German Research Foundation) Project ID 273811115 ‘SFB 1225 ISOQUANT'.
T.V.Z. thanks Christian Kokail and Peter Zoller for valuable discussions.
T.V.Z.'s work is supported by the Simons Collaboration on Ultra-Quantum Matter, which is a grant from the Simons Foundation (651440, P.Z.).

%---------------------------------------------------------------------------------------------------------------------------
% 		REFERENCES
%---------------------------------------------------------------------------------------------------------------------------    
\bibliographystyle{apsrev4-2} 
\bibliography{references}

\section{Supplemental Material}

\subsection{Dual formulations \\of $\mathbf{Z}_2^{2+1d}$, $\mathbf{Z}_N^{2+1d}$ and $U(1)$.}
Dual formulations of $\mathbf{Z}_2^{2+1}$ are well known, for fixed boundary conditions \cite{horn1979hamiltonian} and for an infinite system without a boundary \cite{wegner1971duality}. 

Here, we give exact dual formulations for periodic (PBC) and open boundary (OBC) conditions, as well as 
for systems with `virtual' entanglement cuts. This allows us to compute the entanglement spectrum and entanglement entropy 
of $\mathbf{Z}_2^{2+1d}$ LGT by embedding it into a larger unphysical HS along those boundaries.

\textit{Dual formulation on a periodic lattice.} Considering a periodic lattice with $N_x\times N_y$ sites, where $n_i \in 0,\dots N_i-1$ and $\mathbf{n}\equiv (n_x,n_y)$, the original  $\mathbf{Z}_2^{2+1d}$ gauge variant
electric field variables $\sigma^x_{\mathbf{n},i}$ are written in terms of the dual variables $\mu^x_{\mathbf{x}}$ and  $V_x$, $V_y$
\begin{align}\label{eq:defsigmaPBC1}
\sigma^x_{\mathbf{n},x} = \begin{cases}
\mu^x_{\mathbf{n}} \mu^x_{\mathbf{n}-\hat{y}} & \text{if}\; n_y>0 \;\&\; \mathbf{n}\neq (N_x-1,1)\,,\\
\mu^x_{\mathbf{n}} \mu^x_{\mathbf{n}-\hat{y}}  V_y  & \text{if}\; n_y=0 \;\&\; \mathbf{n}\neq (N_x-1,0)\,,\\
\mu^x_{(N_x-1,N_y-1)}V_xV_y & \text{if}\;  \mathbf{n}=(N_x-1,0)\,,\\
\mu^x_{(N_x-1,1)}V_x &\text{if}\;  \mathbf{n}=(N_x-1,1)\,,
\end{cases}
\end{align}
and
\begin{align}\label{eq:defsigmaPBC2}
\sigma^x_{\mathbf{n},y}=
\begin{cases}
\mu^x_{\mathbf{n}} \mu^x_{\mathbf{n}-\hat{x}}& \text{if}\; n_x>0 \;\&\; \mathbf{n}\neq (N_x-1,0)\,,\\
\mu^x_{\mathbf{n}} \mu^x_{\mathbf{n}-\hat{x}}V_x & \text{if}\; n_x=0 \;\&\; \mathbf{n}\neq (0,0)\,,\\
\mu^x_{(N_x-2,0)}V_x &\text{if}\; \mathbf{n}=(N_x-1,0)\,,\\
\mu^x_{\mathbf{n}} &\text{if}\; \mathbf{n}=(0,0)\,.
\end{cases}
\end{align}
$V_x$ and $V_y$ are independent d.o.f.'s in the dual formulation. The  $N_x \times N_y -1$ independent dual magnetic variables are defined as
\begin{align}
\mu^z_\mathbf{n}=\begin{cases}\label{eq:defmagnvars}
\sigma_{\mathbf{n},x}^z\sigma_{\mathbf{n}\hat{x},y}^z\sigma_{\mathbf{n}\hat{y},x}^z \sigma_{\mathbf{n},y}^z & \text{if}\; \mathbf{n}\neq (N_x-1,0)\,,\\
\prod_{\mathbf{n}'\neq (N_x-1,0)} \mu^z_{\mathbf{n}'}\ & \text{if}\; \mathbf{n}= (N_x-1,0)\,,
\end{cases}
\end{align} 
Just as the original variables,  $\mu^{x/z}_\mathbf{n}$ are spin operators (on the dual lattice); the $N_x \times N_y -1$ independent Gauss laws are eliminated. 
Because the change of variables is canonical, the $Z_2$ algebra remains unchanged $\mu^x_{\mathbf{n}} \mu^z_{\mathbf{n}} \mu^x_{\mathbf{n}} = - \mu^z_{\mathbf{n}}$.
The dual Hamiltonian is
\begin{align}
H=&-\epsilon \sum_{\mathbf{n}} (\sigma^x_{\mathbf{n},x} + \sigma^x_{\mathbf{n},y})
\nonumber\\
-&\sum_{\mathbf{n}\neq (N_x-1,0)} \mu^z_{\mathbf{n}} -  \prod_{\mathbf{n}\neq (N_x-1,0)} \mu^z_{\mathbf{n}}
\end{align}
where $\sigma^x_{\mathbf{n},x}$ and $\sigma^x_{\mathbf{n},y}$ are given by \Eqs{eq:defsigmaPBC1}{eq:defsigmaPBC2}, respectively.

\textit{Dual formulation with open (electric) boundary periodic conditions in the x-direction, and periodic boundary conditions in the y-direction.} 
We consider the case of periodic boundary conditions in the y-direction and open (electric)~\cite{radivcevic2016entanglement,lin2020comments} boundary conditions in x, corresponding to a finite cylinder. Boundary conditions are such that the electric flux into the system is accessible through Gauss law on the boundary, i.e.
$\sigma^x_{(-1,n_y),x} = \sigma^x_{(0,n_y),x}\sigma^x_{(0,n_y),y}\sigma^x_{(0,n_y-1),y}$ and similar at $N_x-1$. Because the corresponding
magnetic pla\-quet\-te terms are not, $\sigma^x_{(-1,n_y),x} $ and $\sigma^x_{(N_x-1,n_y),x}$  become part of the center of the algebra of $A$. 

While the role of $V_y$ is unchanged compared to the case 
with periodic boundary conditions, note that there is no $V_x$ winding around the $x$ direction. Instead, we can define an operator $\tilde{V}_x \equiv \prod_{(\mathbf{n},i \in \tilde{\mathcal{C}})} \sigma^x_{\mathbf{n},i} $ where $\tilde{\mathcal{C}}$
is a path connecting both boundaries. In the dual formulation $\tilde{V}_x = \sigma^x_{ {\scriptscriptstyle(0,n_y),y  }} \mu^x_{(0,n_y)} \sigma^x_{ {  \scriptscriptstyle (N_x-1,n_y'),y   }} \mu^x_{(N_x-1,n_y')} $ with $n_y,n_y'$ arbitrary.
One can show that $\tilde{V}_x$ commutes with all operators in $A$, in particular with the plaquette operators, and is therefore also an element of the center.

Gauss laws are eliminated, except on the boundary where we work with gauge-variant operators  $\sigma^x_{\mathbf{n},y}$ (for $n_x =0$ 
and $n_x=N_x-1$).  The electric field operators are in the bulk, 
\begin{align}\label{eq:defsigmaPBC1virtual}
\sigma^x_{\mathbf{n},y} =
\begin{cases}
\mu^x_{\mathbf{n}} &\text{if}\; \mathbf{n}=(0,0)\,,\\
\mu^x_{\mathbf{n}} \mu^x_{\mathbf{n}-\hat{x}}&\text{else, i.e. for } 0<n_x<N_x-1 \,,
\end{cases}
\end{align}
while on the two open boundaries $\sigma^x_{\scriptscriptstyle (0,n_y),y }$ and $\sigma^x_{\scriptscriptstyle (N_x-1,n_y),y }$ are not replaced by dual variables.
Likewise, the electric field variables in x direction are given as follows
\begin{align}\label{eq:defsigmaPBC2virtual}
\sigma^x_{\mathbf{n},x}
=\begin{cases}
\mu^x_{\mathbf{n}} \mu^x_{\mathbf{n}-\hat{y}}&\text{if}\;  n_y >0\,,\\
\mu^x_{\mathbf{n}} \mu^x_{\mathbf{n}-\hat{y}}V_y&\text{if}\;  n_y =0\,,
\end{cases}
\end{align}
for $0\le n_x <N_x-1$.
The electric flux through the two boundaries at $n_x =-1$ 
and $n_x=N_x-1$ is given by
\begin{align}\label{eq:defsigmaPBC3virtual}
&\sigma^x_{{\scriptscriptstyle(-1,n_y)},x}=
\nonumber\\
&\quad \begin{cases}
\sigma^x_{{\scriptscriptstyle(0,N_y-1)},y}\; \mu^x_{{\scriptscriptstyle(0,N_y-1)}} V_y  & \text{if}\; n_y=0\,,\\
\sigma^x_{{\scriptscriptstyle(0,1)},y}\; \mu^x_{\scriptscriptstyle(0,1)} & \text{if}\; n_y=1\,,\\
\sigma^x_{{\scriptscriptstyle(0,n_y)},y}\sigma^x_{{\scriptscriptstyle(0,n_y-1)},y} \; \mu^x_{\scriptscriptstyle(0,n_y)} \mu^x_{\scriptscriptstyle(0,n_y-1) } & \text{if}\; n_y\ge2\,,
\end{cases}
\end{align}
on the left boundary and
\begin{align}\label{eq:defsigmaPBC4}
& \sigma^x_{{\scriptscriptstyle(N_x-1,n_y)},x}=
\nonumber\\
&\quad\begin{cases}
\sigma^x_{{\scriptscriptstyle(N_x-1,0)},y}\sigma^x_{{\scriptscriptstyle(N_x-1,N_y-1)},y} & {} \\
 \quad\times \mu^x_{{\scriptscriptstyle(N_x-2,0)}} \mu^x_{\scriptscriptstyle(N_x-2,N_y-1) }V_y & \text{if}\; n_y=0\,,\\
\sigma^x_{{\scriptscriptstyle(N_x-1,n_y)},y}\sigma^x_{{\scriptscriptstyle(N_x-1,n_y-1)},y} & {}\\
\quad\times  \mu^x_{\scriptscriptstyle(N_x-2,n_y)} \mu^x_{\scriptscriptstyle(N_x-2,n_y-1) }& \text{if}\; n_y\ge 1\,,
\end{cases}
\end{align}
and the right.
We retain gauge variant magnetic variables near the boundaries, and the magnetic variables can be written as
(for $n_x\in [0,N_x-2]$, $n_y \in [0,N_y-1]$)
\begin{align}\label{eq:defsigmaPBC5}
\mu^z_{\mathbf{n}}=
\begin{cases}
\sigma^z_{\mathbf{n},x} \sigma^z_{\mathbf{n}+\hat{x},y} \sigma^z_{\mathbf{n}+\hat{y},x} &\text{if}\;  n_x =0 \; \&\; n_y >0 \,,\\
\sigma^z_{\mathbf{n},x} \sigma^z_{\mathbf{n}+\hat{y},x} \sigma^z_{\mathbf{n},y} &\text{if}\;  n_x =N_x-2  \,,\\
\sigma_{\mathbf{n},x}^z \sigma_{\mathbf{n}+\hat{x},y}^z\sigma_{\mathbf{n}\hat{y},x}^z \sigma_{\mathbf{n},y}^z  &\text{elsewhere}\,.
\end{cases}
\end{align}
Finally, the dual Hamiltonian is given by,
\begin{align}
H=&- \sum_{n_x=1}^{N_x-3} \sum_{n_y=0}^{N_y-1} \mu^z_{\mathbf{n}}  - \sum_{n_y=1}^{N_y-1} \mu^z_{(0,n_y) }\sigma^z_{(0,n_y),y}
- \mu^{z}_{(0,0)} 
\nonumber\\
&-\sum_{n_y=0}^{N_y-1} \mu^z_{(N_x-2,n_y) }\sigma^z_{(N_x-1,n_y),y}
\nonumber\\
&-\epsilon \sum_{\mathbf{n}} (\sigma^x_{\mathbf{n},x} + \sigma^x_{\mathbf{n},y})\,.
\end{align}
where the last terms, which include $\sigma^x_{(-1,n_y),x} $ and $\sigma^x_{(N_x-1,n_y),x}$, are given upon insertion of \Eqs{eq:defsigmaPBC1}{eq:defsigmaPBC5}.

\textit{Dual formulation with virtual boundaries.} We consider now a torus with entanglement cuts as shown in the main text, with $ (N_x^A+N_y^B)\times N_y$ lattice sites, separated into subsystems $A$ and $B$.
The boundary of $A$ is chosen
as open electric boundary conditions as above. Again the electric flux into $A$ and $\tilde{V}_x$ are part of the center of the algebra of $A$. (Apart from Gauss on the boundary, they 
are the only elements of the center because Gauss law is eliminated in the bulk.)

In this formulation, the Hamiltonian is
\begin{align}
H = H_A + H_B + H_{AB}
\end{align}
where
\begin{align}
H_A \equiv &- \mu^z_{(0,0)} -\sum_{n_x=1}^{N_x^A-3} \sum_{n_y=0}^{N_y-1 }\mu^z_{\mathbf{n}} - \sum_{n_y=1}^{N_y-1} \mu^z_{(0,n_y)} 
\sigma^z_{(0,n_y),y}
\nonumber\\
&  -\sum_{n_y=0}^{N_y-1} \mu^z_{(N_x^A-2,n_y)} \sigma^z_{(N_x^A-1,n_y),y} - \epsilon \mu^x_{(0,0)} 
\nonumber\\
&
-\epsilon\sum_{n_y=1}^{N_y-1}\sigma^x_{(0,n_y),y}
  - \epsilon \sum_{n_y=0}^{N_y-1} \sigma^x_{(N_x^A-1,n_y),y}
  \nonumber\\
&
 -\epsilon \sum_{n_x=1}^{N^A_x-2}\sum_{n_y=0}^{N_y-1} \mu^x_{\mathbf{n}}\mu^x_{\mathbf{n}-\hat{x}}
-\epsilon \sum_{n_x=0}^{N^A_x-2}\sum_{n_y=1}^{N_y-1} \mu^x_{\mathbf{n}}\mu^x_{\mathbf{n}-\hat{y}}
  \nonumber\\
&
  -\epsilon \sum_{n_x=0}^{N_x^A-2}\mu^x_{(n_x,0)}\mu^x_{(n_x,N_y-1)}V_y\,,
\end{align}
includes operators in $A$. Further
\begin{align}
H_B\equiv& -\sum_{n_x=N_x^A}^{N_x-2} \sum_{n_y=0}^{N_y-1} \mu^z_{\mathbf{n}}
- \epsilon V_y \mu^x_{(N_x-1,N_y-1)}
  \nonumber\\
&
 - \epsilon \sum_{n_y=2}^{N_y-1} \mu^z_{(N_x-1,n_y)}\mu^z_{(N_x-1,n_y-1)} - \epsilon \mu^x_{(N_x-1,1)}
\nonumber\\
&  -\epsilon \sum_{n_y=1}^{N_y-1} \mu^x_{(N^A_x-1,n_y)} \mu^x_{(N^A_x-1,n_y-1)}
  \nonumber\\
&- \epsilon \mu^x_{(N^A_x-1,0)} \mu^x_{(N^A_x-1,N_y-1)}V_y
- \epsilon \sum_{N_x^A}^{N_x-2}\sum_{n_y=1}^{N_y-1} \mu^x_{\mathbf{n}}\mu^x_{\mathbf{n}-\hat{y}}
  \nonumber\\
&-\epsilon \sum_{n_x=N_x^A}^{N_x-2}\mu^x_{\scriptscriptstyle (n_x,0)}\mu^x_{\scriptscriptstyle (n_x,N_y-1)} V_y - \epsilon\mu^x_{(N_x-2,0)}
\nonumber\\
&  -\epsilon \sum_{n_x=N_x^A}^{N_x-1} \sum_{n_y=0}^{N_y}\big|_{\mathbf{n}\neq (N_x-1,0)} \mu^x_{\mathbf{n}}\mu^x_{\mathbf{n}-\hat{x}} \,,
\end{align}
includes operators in $B$, while terms with operators in both $A$ and $B$ are
\begin{align}\label{eq:HAB}
H_{AB}\equiv& - \sum_{n_y=0}^{N_y-1} \mu^z_{(N_x^a-1,n_y)} \sigma^z_{(N_x^A-1,n_y)} 
\nonumber\\
& 
-\sum_{n_y=1}^{N_y-1} \mu^z_{(N_x-1,n_y)} 
\sigma^z_{(0,n_y),y} 
\nonumber\\
& 
- \text{(prod.  all plaquettes A \& B)}
\end{align}
where `prod. all plaquettes A \& B`  means that the plaquette term at  $\mathbf{n}=(N_x-1,0)$ is given by the product of all magnetic variables elsewhere, i.e. as in \Eq{eq:defmagnvars}. We have checked that this is a canonical transformation, preserving the canonical commutation relations, and numerically compared the spectra of original and dual theory.

Dual formulations of $\mathbf{Z}_N^{2+1}$~\cite{horn1979hamiltonian} with Hamiltonian, 
\begin{align}\label{eq:ZNHamiltonian}
H_N= &- \sum_{\mathbf{n}} Q_{\mathbf{n},x}Q_{\mathbf{n}+\hat{x},y}Q^\dagger_{\mathbf{n}+\hat{y},x}Q^\dagger_{\mathbf{n},y}  + \rm{h.c.}
\nonumber\\
& -\epsilon\sum_{\mathbf{n},i=x,y} (P_{\mathbf{n},i}+P^\dagger_{\mathbf{n},i}) \,,
\end{align}
and algebra $P^\dagger_{\mathbf{n}} Q_{\mathbf{n}} P_{\mathbf{n}} = e^{ i\frac{2\pi}{N}}Q_{\mathbf{n}}$,  $P^N_{\mathbf{n}} = Q^N_{\mathbf{n}} =1$, Gauss law $P_{\mathbf{n},x} P^\dagger_{\mathbf{n}-\hat{x},x} P_{\mathbf{n},y} P^\dagger_{\mathbf{n}-\hat{y},y}$,
with both virtual and physical boundaries, are straightforward generalizations of the $\mathbf{Z}_N^{2+1}$ case. The difference is that the variables $\mu^x_{\mathbf{n}}, \mu^z_{\mathbf{n}}$ are complex, e.g.
\begin{align}
P_{\mathbf{n},x} = \mu^{x\dagger}_{\mathbf{n}}   \mu^{x}_{\mathbf{n}-\hat{y}}  \,,\qquad P_{\mathbf{n},y} = \mu^{x\dagger}_{\mathbf{n}}   \mu^{x}_{\mathbf{n}-\hat{x}}  
\end{align}
and
\begin{align}
\mu^z_{\mathbf{n}} = Q_{\mathbf{n},x}Q_{\mathbf{n}+\hat{x},y}Q^\dagger_{\mathbf{n}+\hat{y},x}Q^\dagger_{\mathbf{n},y}  
\end{align}
in the bulk. One uses combinations of original gauge-variant and dual variables on boundaries, in direct analogy to \Eqs{eq:defsigmaPBC1}{eq:HAB}. The dual variables obey a the local $Z_N$ algebra $\mu^{x\dagger}_{\mathbf{n}} \mu^z_{\mathbf{n}} \mu^x_{\mathbf{n}} = e^{ i\frac{2\pi}{N}} \mu^z_{\mathbf{n}}$
as well as $ (\mu^z_{\mathbf{n}})^N = (\mu^x_{\mathbf{n}})^N =1$. $U(1)$ LGT is obtained in the limit $N\rightarrow \infty$.
%
%
% \subsection{Details of Perturbative Computation}

 \subsection{Topological Order versus Confinement Phase Transition}
 \begin{figure}[h]
\begin{center}
\includegraphics[width=0.31\textwidth]{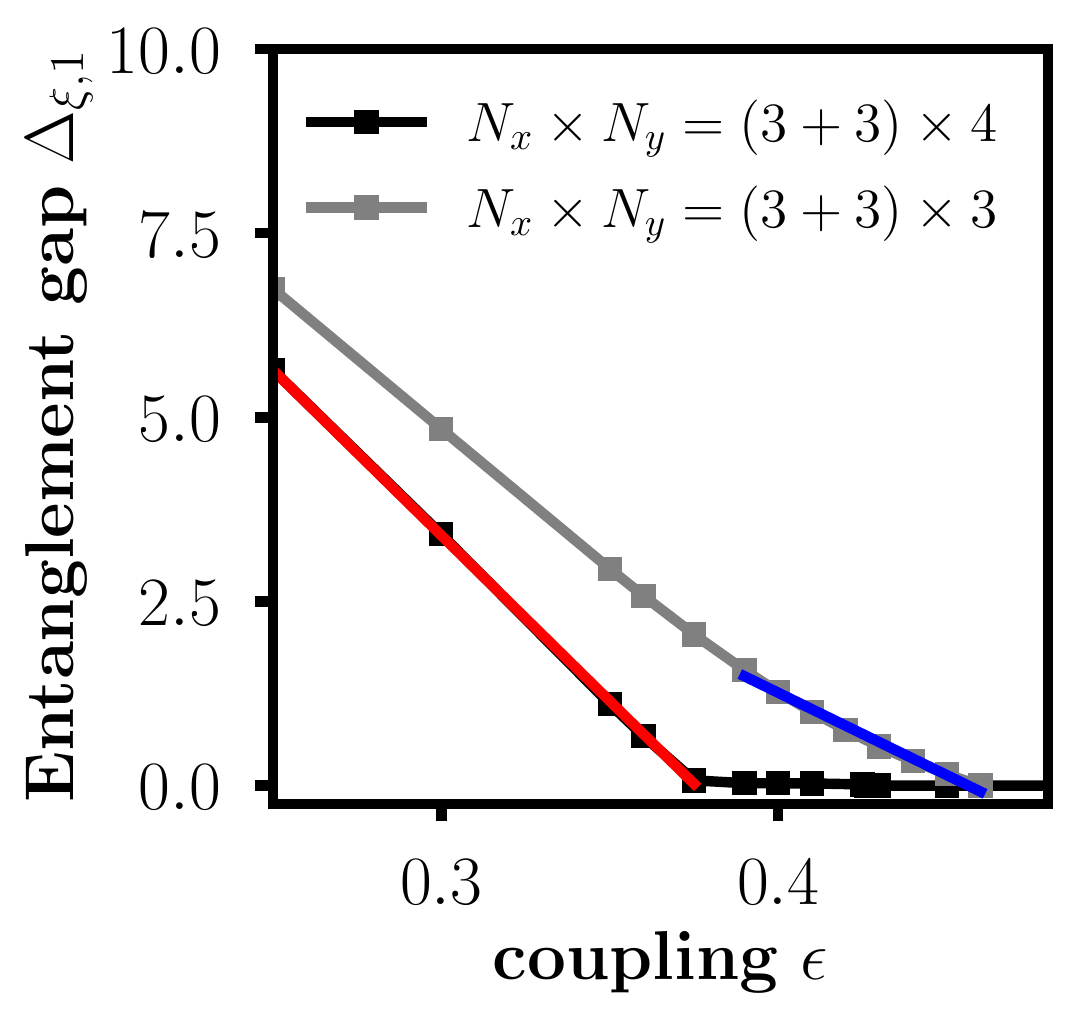}
\caption{Details of gap closing $\Delta_{\xi,{1}}\rightarrow 0$ for $(N_x^A+N_x^B)\times N_y = (3+3)\times 4$ and 
$(N_x^A+N_x^B) \times N_y = (3+3)\times 4$ to determine the position $\epsilon_c$ of the confinement/deconfinement transition.}
\label{fig:ESsupp}
\end{center}
\end{figure} 

\begin{figure}[t]
\begin{center}
\includegraphics[width=0.38\textwidth]{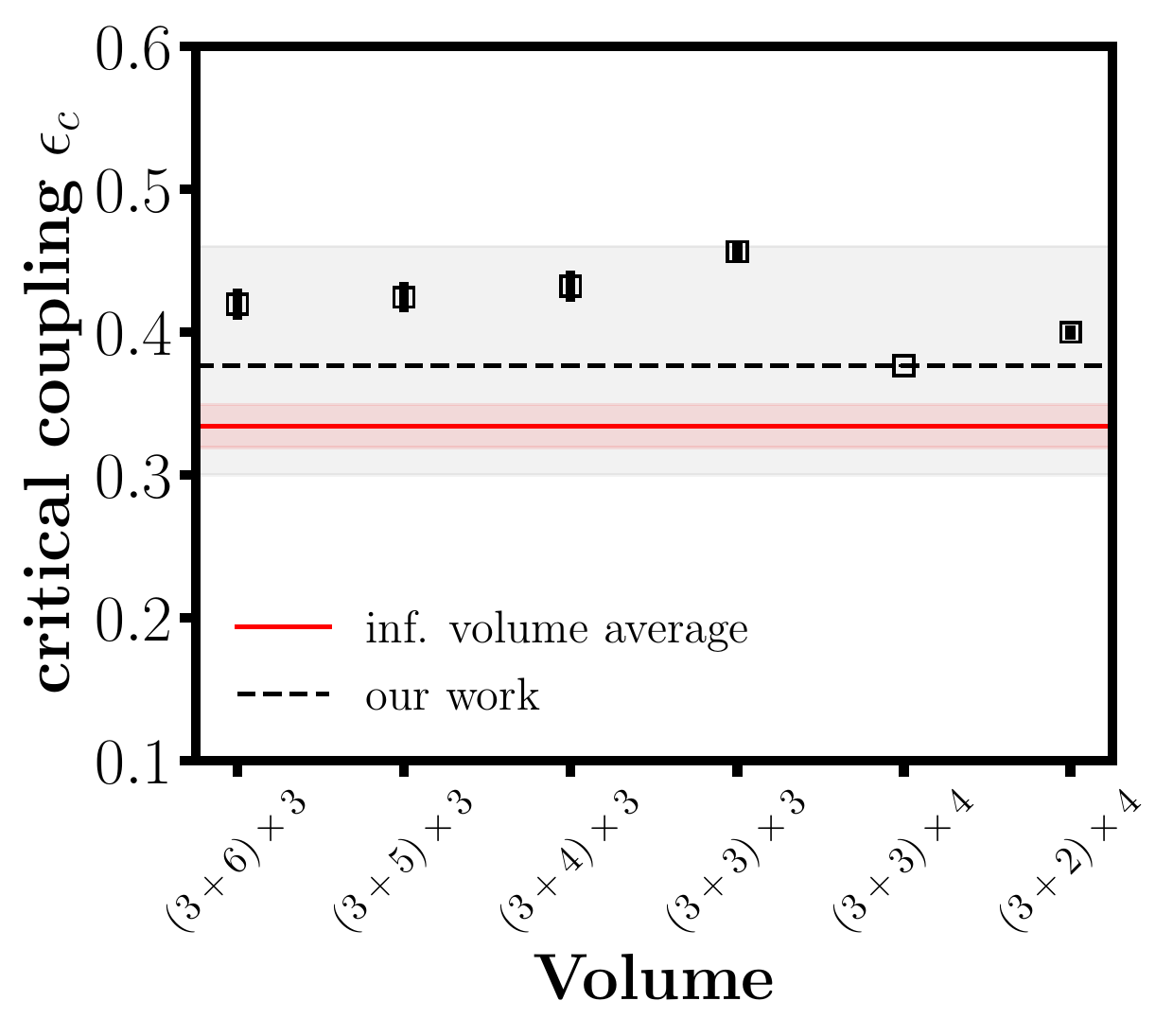}
\caption{Finite-size analysis of the critical coupling $\epsilon_c$, measured as the closing of the Entanglement Gap of the Entanglement Spectrum, for different lattice sizes $(N_x^A+N_x^B)\times N_y$.}
\label{fig:static}
\end{center}
\end{figure}

To determine the position of the confinement/deconfinement transition $\epsilon_c$, we consider the closing of the first entanglement gap $\Delta_{\xi,{1}}\rightarrow 0$, shown in \Fig{fig:ESsupp} for $(N_x^A+N_x^B) \times N_y = (3+3)\times 4$ and 
$(N_x^A+N_x^B) \times N_y = (3+3)\times 4$. We use a linear fit to determine the intercept $\epsilon_c$. We determine the uncertainty of this linear extrapolation by removing one data point and repeating the fit. The uncertainty of this extrapolation is
less than $0.01$ for all lattice sizes and is shown as error bars in \Fig{fig:static}. We use the difference between the lattices as an estimate for the finite size error, resulting, together with the uncertainty of the linear extrapolation, in $\epsilon_c =0.38\pm 0.09$.
While an infinite volume extrapolation is not feasible with exact diagonalization, we note that the infinite volume limit result is well within the uncertainties we give.

We conducted a finite-size analysis, performing additional runs varying the size of the lattice in both $x$- and $y$-directions, with results compactly summarized in \Fig{fig:static}. These results show the robustness of the entanglement gap as an `order parameter', demonstrating that finite volume effects are well under control. We note that in contrast, traditionally, the confinement/deconfinement transition is determined by expectation values of Wilson loop operators. Doing so, distinguishing volume and area law scaling, is very difficult on small lattices.

\subsection{Optimal local approximation \\of the Entanglement Hamiltonian}
Given the exact reduced state $\rho_A$, we wish to obtain an optimal local approximation
$\rho_A \approx \sigma_A$ where
\begin{align}
 \sigma_A \equiv \frac{1}{Z} e^{ -\sum_{\mathbf{n}\in A} \beta_\mathbf{n} h_\mathbf{n} } \;, && Z = \text{Tr} \left[ e^{ -\sum_{\mathbf{n}\in A} \beta_\mathbf{n} h_\mathbf{n} } \right]
\end{align}
where $Z=Z(\{ \beta_{\mathbf{n}}\})$,
with variational parameters $\beta_\mathbf{n}$ and local operators $h_\mathbf{n}$. We obtain $\sigma_A$ by minimizing the relative entropy (Kullback-Leibler divergence)~\cite{kullback1951information}
\begin{align}
S(\rho_A||\sigma_A) &\equiv \text{Tr} \left[ \rho_A \left(\log \rho_A - \log \sigma_A\right) \right] \nonumber\\
&= -S(\rho_A) + \log Z + \sum_\mathbf{n} \beta_\mathbf{n} \langle h_\mathbf{n} \rangle \;,
\end{align}
where
$\langle h_\mathbf{n} \rangle = \text{Tr} \left[\rho_A h_\mathbf{n} \right]$ are the expectation values in the state $\rho_A$ and $S(\rho_A)$ is the corresponding von Neumann entropy.  Motivated by the BW theorem, in practice we choose all contributions of the subsystem to the Hamiltonian, i.e. all electric and magnetic energy terms fully supported in $A$,
as the operators $h_\mathbf{n}$.
It is well known~\cite{kullback1951information} that the relative entropy is a convex function with $S(\rho_A||\sigma_A)\ge 0$ where equality is obtained if $\rho_A = \sigma_A$. While this enables an efficient minimization, we nevertheless need to evaluate $Z$ several hundreds to thousands of times in order to obtain convergence. Since our numerical simulations are based on exact diagonalization, we are limited to relatively small system sizes which we have slightly increased by incorporating translational invariance of the variational parameters parallel to the entanglement cut.

  \begin{figure*}[ht]
\centering{
\includegraphics[scale=0.29]{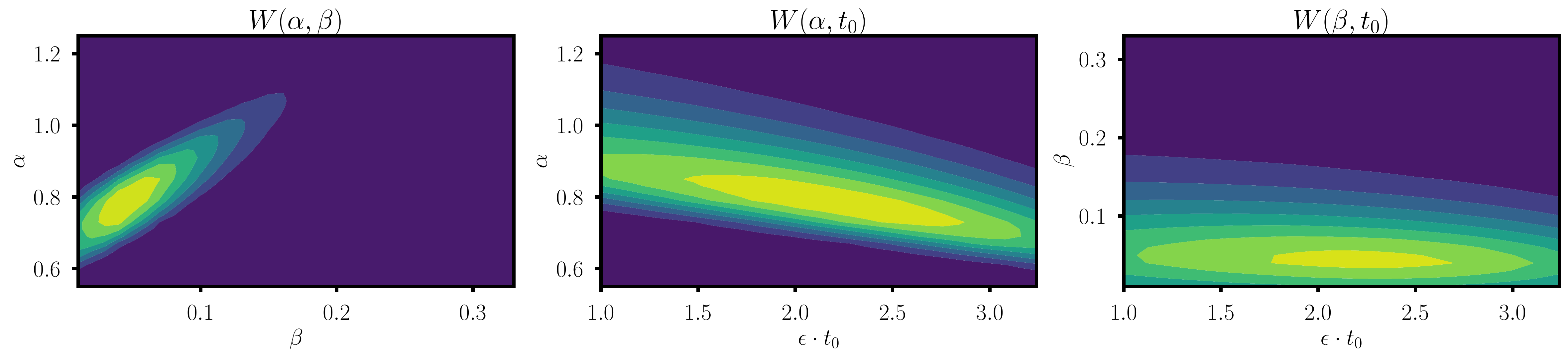}\\
\includegraphics[scale=0.28]{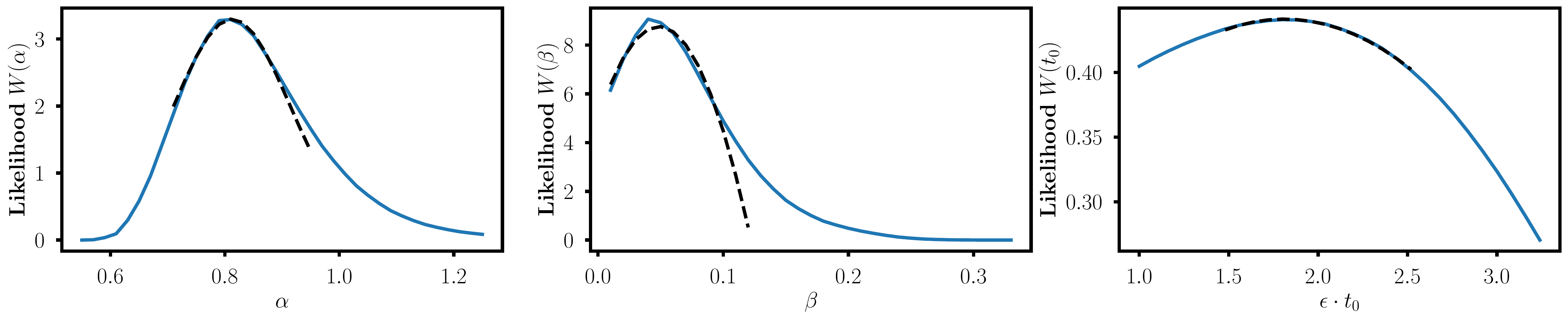}
}
\caption{Top: Partially integrated Likelihood distributions $W(\alpha,\beta)$, $W(\alpha,t_0)$ and $W(\beta,t_0)$. Bottom: Integrated Likelihood distributions $W(\alpha)$, $W(\beta)$ and $W(t_0)$. Dashed lines indicate a Gaussian fit.}\label{fig:statisticalana}
\end{figure*}
 
 \subsection{Level Spacing Statistics\\of the Entanglement Spectrum}
 To extract level spacing statistics of the ES, we first project the reduced density matrix into symmetry sectors. As outlined above, $2^{2 N_y}$ sectors are given by the values of the electric flux into the subsystem
 through both boundaries, i.e. $ \sigma^x_{\scriptscriptstyle(-1,n_y) , x} $ and $ \sigma^x_{\scriptscriptstyle(N^A_x-1,n_y) , x} $. Because of Gauss law on the boundary, 
 these can be written in terms of operators inside system A, e.g.
 \begin{align}
 \sigma^x_{\scriptscriptstyle(-1,n_y) , x} =\sigma^x_{\scriptscriptstyle(0,n_y) , x}  \sigma^x_{\scriptscriptstyle(0,n_y) , y}  \sigma^x_{\scriptscriptstyle(0,n_y-1) ,y} 
 \end{align}
 for the left boundary of $A$. In the dual theory with entanglement cuts  these operators are given as \Eq{eq:defsigmaPBC3virtual} on the left boundary at $n_x=0 $, and on the right boundary at  $n_x = N_x^A-1$
 by \Eq{eq:defsigmaPBC4}. Projectors are simply constructed as
 \begin{align}
 P_{\ua /\da, n_y } = \frac{1}{2} (1\pm \sigma^x_{\scriptscriptstyle(-1,n_y) , x} )
 \end{align}
 and likewise for the other boundary. We label symmetry sectors by a string of symbols $\da\ua \dots \ua$ of length $2^{2N_y}$, starting at $n_y=0$ of the left entanglement cut and ending at $N_y-1$ of the right boundary.
 
 The symmetry related to  $\tilde{V}_x = \prod_{\mathbf{n}\in \tilde{\mathcal{C}} } \sigma^x_{\mathbf{n,i}}$, where $\tilde{\mathcal{C}}$ is a path from one boundary to the other, is projected on by $P = \frac{1}{2}(1\pm \tilde{V}_x)$ and sectors are labelled by $\pm$. Together one of $2^{2N_y+1}$ symmetry sectors is labelled by a string, e.g. $\da\ua \dots \ua + $.

We then compute the gap ratio 
\begin{align}
r_n\equiv \frac{\text{min}(\delta_n,\delta_{n-1})}{\text{max}(\delta_n,\delta_{n-1})}
\end{align}
where $\delta_n\equiv \xi_n -\xi_{n-1} \ge 0$ of the ES $\xi_n$,
between every state $n$  within each symmetry sector and plot the distribution of gap ratios combined from all sectors. We compare
 with a Poissonian distribution,  Gaussian Orthogonal Ensemble (GOE) and  Gaussian Unitary Ensemble (GUE)~\cite{atas2013distribution}
 \begin{align}
& \mathcal{P}^{\rm Poiss.}(r) = \frac{2}{(1+r)^2}\,,\nonumber\\
 & \mathcal{P}^{\rm GOE}(r) =\frac{27}{4} \frac{r+r^2}{(1+r+r^2)^{5/2}}\nonumber\\
& \mathcal{P}^{\rm GUE}(r) =\frac{81}{2} \frac{\sqrt{3}}{\pi}  \frac{(r+r^2)^2}{(1+r+r^2)^{4}}\,.
  \end{align}
  with average $\langle r \rangle^{\rm Poisson} \approx 0.38$, $\langle r \rangle^{\rm GOE} \approx 0.52$ and $\langle r \rangle^{\rm GUE} \approx 0.60$.

For the level spacing ratio we remove the dependence on the mean level density by first unfolding the spectrum $\xi_n \rightarrow \bar{\xi}_n$ as in~\cite{guhr1998random}. We then define $s_n \equiv  \bar{\xi}_n- \bar{\xi}_{n-1}$ and compare with
 \begin{align}
& \mathcal{P}^{\rm Poiss.}(s) = e^{-s}\nonumber\\
 & \mathcal{P}^{\rm GOE}(s) =\frac{\pi}{2} s \, e^{-\frac{\pi}{4} s^2}  \nonumber\\
& \mathcal{P}^{\rm GUE}(s) = \frac{32}{\pi^2}\,s^2 \, e^{-\frac{4}{\pi}s^2}\,.
  \end{align}

 \subsection{Scaling Analysis}
Outlined below are the details of the statistical analysis to determine the scaling coefficients $\alpha,\beta$, closely following \cite{mace2020chiral}. Assuming a scaling form of the Schmidt spectrum of states $n$ in an intermediate time window,
\begin{align}
P(n,t) = \tau^{-\alpha} P( \tau^\beta n)\,,
\end{align}
where $\tau = \epsilon(t- t_0)$ and $t_0 $ is the onset time of self-similarity, we determine $\alpha,\,\beta$ and $t_0$ by the following procedure: First, we define a reference function
\begin{align}
f_{\rm ref}(t, \hat{n} = \tau^\beta_{\rm} n) = \log\{ \tau^{-\alpha}_{\rm ref} P(\tau_{\rm ref}^\beta n ) \}\,.
\end{align}
where $\epsilon \cdot t_{\rm ref} = 6$. We compare this reference function to a number of $N_t$ test times, 
\begin{align}
f_{\rm test}(t, \hat{n} )= \log\{ \tau^{-\alpha}_{\rm test} P(\tau_{\rm test}^\beta n ) \}\,.
\end{align}
where $\epsilon \cdot t_{\rm test} \in\{ 8,12,16,24,30,40,50\}$.  We quantify deviation from ideal scaling in terms of
\begin{align}\label{eq:chis}
\chi^2(\alpha,\beta,t_0) = \frac{1}{N_t} \sum_{t \in t_{\rm test}} \frac{\int  \frac{d\hat{n}}{\hat{n}} (f_{\rm ref}(\hat{n}) - f_{\rm test}(\hat{n})  )^2 }{\int  \frac{d\hat{n}}{\hat{n}} (f_{\rm ref}(\hat{n}))^2}\,.
\end{align}
and restrict $n$ to the scaling window $n \in[130,1300 ]$ which we vary to determine the error of our fit. We checked that a different choice of test times within the scaling regime 
does not change the outcome of our fit.
Note that the spectrum $\hat{n}$ is discrete, and thus the integral should be replaced by a sum, but the same equation holds in the limit of a continuous spectrum.

The likelihood of a given set of parameter to reproduce the ideal rescaling function is given by
\begin{align}
W(\alpha,\beta,t_0) = \frac{1}{\mathcal{N}} \exp \Big( - \chi^2(\alpha,\beta,t_0) / \chi^2_{\rm min} \Big)
\end{align}
where $\mathcal{N}$ is a normalization, and $\chi^2_{\rm min}$ is the minimum of \Eq{eq:chis}, $\chi_{\rm min} = 4\cdot 10^{-5}$, which is obtained for
$\alpha \approx 0.75$, $\beta \approx 0.04$ and $\epsilon \cdot t_0 \approx 2.0$, shown in  the main text. 

\Fig{fig:statisticalana} (top) shows the reduced likelihood distributions $W(\alpha,\beta)$, $W(\alpha,t_0)$ and $W(\beta,t_0)$ obtained by integrating one of three variables, i.e. $W(\alpha,\beta, )  = \int dt_0W(\alpha,\beta,t_0) $. \Fig{fig:statisticalana} (bottom) shows the
integrated distributions $W(\alpha)$, $W(\beta)$ and $W(t_0)$, and the values we quote in the main text are determined from these distributions,
$$
\alpha = 0.8 \pm 0.2\,  \quad \beta = 0.0 \pm 0.1\,,
$$
and $\epsilon \cdot t_0 = 1.8\pm 0.5$. These were determined by fitting \Fig{fig:statisticalana} (bottom) locally with Gaussians. \begin{figure}[t]
\begin{center}
\includegraphics[width=0.38\textwidth]{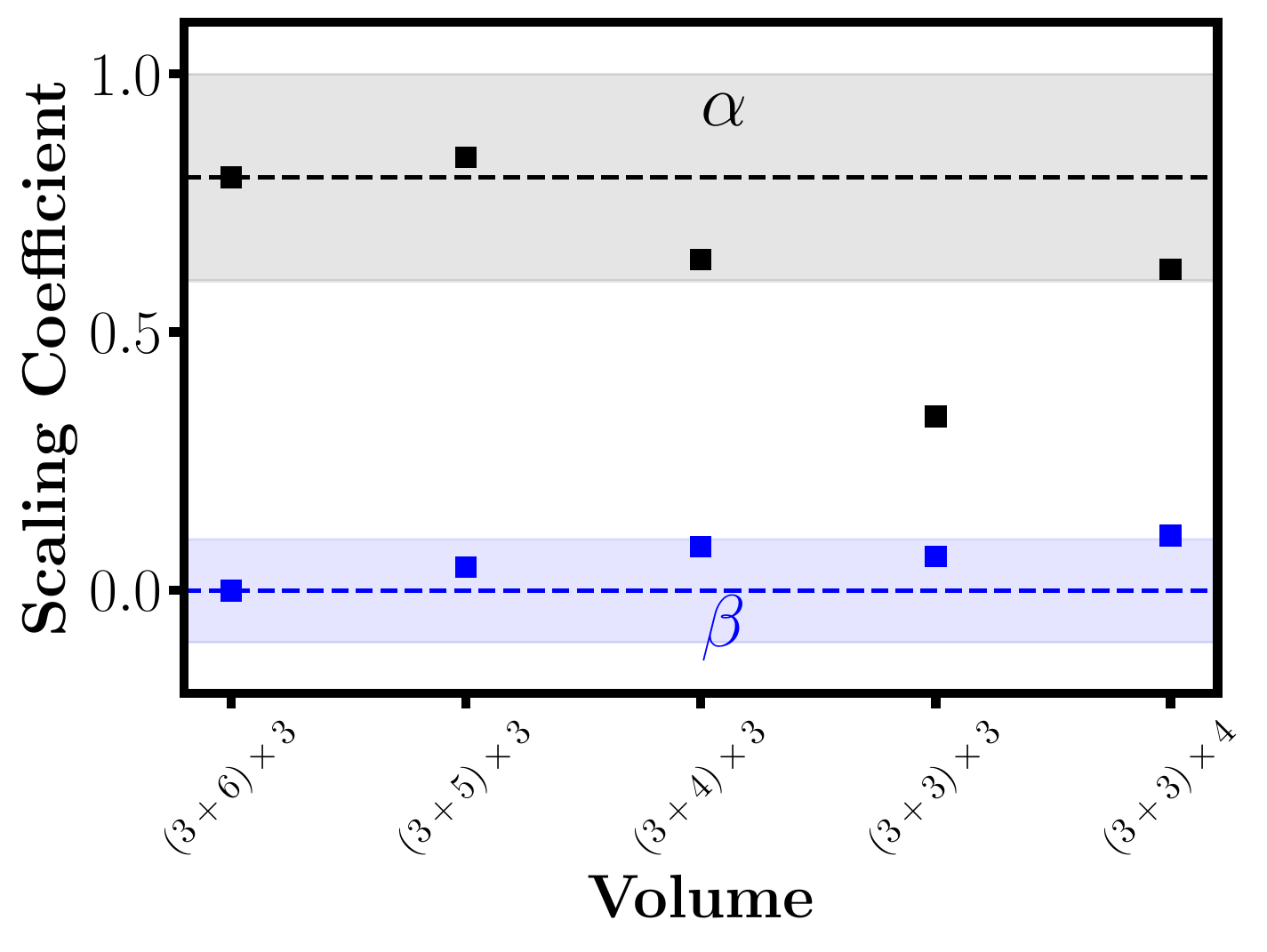}
\caption{Scaling coefficients $\alpha$ and $\beta$ extracted from the Entanglement Spectrum of various lattice sizes.}
\label{fig:scaling}
\end{center}
\end{figure}%

We performed additional runs to estimate finite-volume errors in determining $\alpha$ and $\beta$, with results compactly summarized in \Fig{fig:scaling}. The errors quoted here 
are determined from this analysis, and also include the error from varying the fit range of the statistical analysis. While finite volume effects are somewhat larger than for static quantities,  \Fig{fig:scaling} shows that they are under control, validating our results.

Resolving the scaling regime with higher precision, in particular where it extends into the low lying part of the ES, would require lattices with larger subsystem size of $A$ (and consequently also $V_B \gg V_A$, or at least $V_B > V_A$,  otherwise the Schmidt spectrum of A is cut off). This is challenging given the exponentially larger cost when using exact diagonalization.%, here system $A$ has $N^A_x \times N_y = 3\times 3$ lattice sites. 

\end{document}